\documentclass[11 pt]{article}
\pdfoutput=1

\usepackage[utf8]{inputenc}
\usepackage{amsmath}
\usepackage{amssymb}
\usepackage[margin=1in]{geometry}
\usepackage{graphicx}
\usepackage{caption}

\usepackage{jheppub}
\usepackage{amsthm}
\usepackage{dsfont}

\usepackage{subfigure}

\usepackage{tikz}
\usetikzlibrary{calc}
\usetikzlibrary{decorations.text}
\usetikzlibrary{decorations.pathreplacing,decorations.pathmorphing}
\usetikzlibrary{shapes}
\usetikzlibrary{patterns}

\usepackage{hyperref}

\definecolor{3dcolor}{rgb}{0.96,0.89,0.76}
\definecolor{4dcolor}{rgb}{0.812,0.851,0.914}

\def\CC{\ensuremath{\mathds C}}
\def\RR{\ensuremath{\mathds R}}
\def\ZZ{\ensuremath{\mathds Z}}

\def\Im{\mathop{\rm Im}}
\def\Re{\mathop{\rm Re}}

\preprint{LCTP-23-19}

\title{Binary AdS black holes coupled to a bath in Type IIB}

\author[a]{Evan Deddo,}

\author[a,b]{Leopoldo A.~Pando Zayas,}

\author[c]{Christoph F.~Uhlemann}

\emailAdd{evdedd@umich.edu, lpandoz@umich.edu, christoph.uhlemann@vub.be}

\affiliation[a]{Leinweber Center for Theoretical Physics, 
University of Michigan, Ann Arbor, MI 48109, USA}
\affiliation[b]{The Abdus Salam International Centre for Theoretical Physics, 34014 Trieste, Italy}
\affiliation[c]{Theoretische Natuurkunde, Vrije Universiteit Brussel, Pleinlaan 2, B-1050 Brussels, Belgium}

\abstract{We construct Type IIB string theory setups which, via double holography, realize two gravitational systems in separate AdS spaces which interact with each other and with a non-gravitational bath. We employ top-down string theory solutions with concrete field theory duals in the form of 4d $\mathcal N=4$ SYM BCFTs and a first-principles notion of double holography. The setups are used to realize pairs of `near' and `far' black holes from the perspective of the bath, which exchange Hawking radiation with each other and radiate into the bath. We identify three phases for the entropy in the bath characterized as no island, partial island and full island, and discuss the entropy curves. The setups differ from the black hole binaries observed in gravitational wave experiments but may  capture certain aspects.}

\keywords{}

\date{\today}

\begin{document}

\maketitle

\section{Introduction}
Recent progress on the black hole information paradox, driven by an improved understanding of entropy computations, has led to information accounting which is consistent with unitarity, i.e.\ Page curves \cite{Penington:2019npb,Almheiri:2019psf,Almheiri:2019hni,Almheiri:2019yqk,Almheiri:2019psy,Penington:2019kki,Almheiri:2019qdq,Sully:2020pza,Chen:2020uac,Chen:2020hmv} (for reviews see \cite{Almheiri:2020cfm,Raju:2020smc}). Concrete calculations have employed black holes coupled to a bath, mostly in simplified 2d models or in bottom-up braneworld models with no clear UV completion. While higher-dimensional black holes near extremality can be connected to 2d models, general higher-dimensional black holes are genuinely harder. A setting where higher-dimensional black holes coupled to a QFT bath have been consistently embedded in string theory are the models engineered in \cite{Uhlemann:2021nhu} for the computation of Page curves for $\rm AdS_4$ black holes coupled to 4d $\mathcal N=4$ SYM as bath.

In this work we generalize the top-down setups in \cite{Uhlemann:2021nhu} motivated by the following question: What entropy curve should an observer see who is collecting radiation from two interacting black holes in a region of a QFT bath, to which the two black holes are coupled? Our setups differ in crucial ways from the binary black holes observed in gravitational wave experiments, but they may be seen as a small step towards studying information transfer out of real-world black hole binaries.
In our setups we will have two black holes in separate $\rm AdS_4$ spaces, which are coupled to each other at their conformal boundaries. The mechanism preventing the two black holes from merging is the inherent gravitational pull in AdS, which acts as a confining box (rather than angular momentum). The two black holes are further coupled to a non-gravitational bath at the same temperature as the black holes.\footnote{The coupling to the bath generally introduces a graviton mass in the AdS gravity theories \cite{Geng:2020qvw}. The mass is determined by the details of the coupling, which we spell out explicitly. The mass can be tuned and there is no vDVZ discontinuity in AdS \cite{Porrati:2000cp}, but for non-zero coupling it is part of the models. See also \cite{Bachas:2018zmb,Bachas:2017rch}.}
All three sectors exchange radiation so as to realize a steady state.

Working in top-down constructions has a crucial advantage: The Page curve discussions in higher dimensions rely on the concept double holography. This is the idea that boundary CFTs (BCFTs) can have, besides a conventional holographic description geometrizing all BCFT degrees of freedom, an additional ``intermediate" description, which emerges when only the boundary degrees of freedom of the BCFT are geometrized.
This leads to a gravitational theory coupled to the leftover ambient field theory degrees of freedom as a ``bath".
The concept was first formulated in bottom-up holographic duals for BCFTs with an ``end of the world" brane terminating spacetime \cite{Karch:2000gx,Karch:2000ct,Takayanagi:2011zk}. However, in the bottom-up models there is no precise definition of the BCFTs or how the boundary and ambient CFT degrees of freedom are coupled, and in general no precise holographic dictionary for the intermediate description.\footnote{A bottom-up dictionary for the near-critical limit was proposed in \cite{Neuenfeld:2021wbl}. The bottom-up notion of double holography in general leads to inconsistencies between the causal structures, discussed in \cite{Omiya:2021olc,Neuenfeld:2023svs}.}
In the top-down models, on the other hand, the ideas behind double holography can be made precise \cite{Karch:2022rvr}, with concrete BCFTs and coupling between gravity and bath, and a precise holographic dictionary for the intermediate description.

The top-down constructions are based in Type IIB string theory, where a broad class of 4d BCFTs with 3d $\mathcal N=4$ defect superconformal symmetry can be realized by D3-branes ending on D5-branes and NS5-branes \cite{Gaiotto:2008sa,Gaiotto:2008sd}. In field theory terms this leads to 3d $\mathcal N=4$ SCFTs called $T_\rho^\sigma[SU(N)]$ on the boundary of a 4d half space on which a 4d $\mathcal N=4$ SYM theory resides. The coupling between them is determined by the concrete brane configuration.
The 3d $T_\rho^\sigma[SU(N)]$ SCFTs are a broad class of boundary degrees of freedom and give ample room for holographic model building. They can be isolated out of the BCFT and dualized holographically either on their own, or together with the 4d ambient CFT, as described in \cite{Karch:2022rvr}, using the solutions of \cite{DHoker:2007zhm,DHoker:2007hhe,Aharony:2011yc,Assel:2011xz}.

The setups were used to realize Type IIB string theory versions of double holography for BCFTs in \cite{Uhlemann:2021nhu}, with a precise holographic dictionary given in \cite{Karch:2022rvr}. 
The basic idea for double holography may be summarized schematically as follows,
\begin{equation}\label{eq:double-hol-BCFT}
\begin{tikzpicture}
	\node at (-0.3,0) {\sf 4d BCFT \ $=$ \ 3d defect d.o.f.\ \ $+$ \ 4d ambient d.o.f.};
	\draw [decorate,decoration = {brace,amplitude=6pt}] (-2.1,0.4) --  (3.8,0.4);
	\node at (1.8,1.0) {\sf\small QFT$_d$ $=$ gravity$_{d+1}$ for full 4d BCFT $\rightarrow$ asymptotically locally $\rm AdS_5$ gravity};
	
	\draw [decorate,decoration = {brace,amplitude=5pt}] (0.3,-0.4) -- (-2.1,-0.4);
	\node at (1.2,-1.0) {\sf\small QFT$_3$ $=$ gravity$_{4}$ $\rightarrow$ $\rm AdS_4$ gravity $+$ 4d ambient CFT};
\end{tikzpicture}
\end{equation}
The field theory is at the center. The full holographic description, which geometrizes all field theory degrees of freedom using AdS/CFT dualities, is on top. The intermediate level description, which is the setting for a black hole coupled to a bath, is obtained by geometrizing only the 3d boundary degrees of freedom and shown at the bottom. The calculations happen at the top, but the results are interpreted in the intermediate description at the bottom.\footnote{We note that the choice of the 3d defect d.o.f.\ in the BCFT which are geometrized separately for the intermediate picture need not be unique; different decompositions can lead to different intermediate pictures.}
The calculations \cite{Uhlemann:2021nhu} for an eternal black hole exchanging radiation with a bath at the same temperature show that the radiation entropy either follows the Page curve for eternal black holes, which saturates after an initial growth period, or is altogether flat, depending on the size of the radiation region. Both options are consistent with unitarity. 

Similar setups were also used in \cite{Uhlemann:2021nhu} to give a string theory realization of the bottom-up wedge holography proposal of \cite{Akal:2020wfl}, which is a variation of double holography.
The concept behind wedge holography can be summarized in a similar way. One starts from a 3d CFT which can be decomposed into two subsectors which are coupled to each other. If the full CFT as well as the two subsectors all admit separate holographic duals, one has a choice whether to geometrize the full CFT or the subsectors separately. 
This idea was used in bottom-up models in \cite{Geng:2020fxl} to study black holes coupled to a gravitating bath.
Schematically,
\begin{equation}\label{eq:double-hol-wedge}
	\begin{tikzpicture}
		\node at (-1.0,0) {\sf 3d SCFT \ \ \ $=$ \ \ \ 3d SCFT$_1$ \ \ \ $+$ \ \ \ 3d SCFT$_2$};
		\draw [decorate,decoration = {brace,amplitude=6pt}] (-2.1,0.4) --  (2.8,0.4);
		\node at (0.8,1.0) {\sf\small QFT$_3$ $=$ gravity$_{4}$ for full 3d CFT $\rightarrow$ $\rm AdS_4$ gravity};
		
		\draw [decorate,decoration = {brace,amplitude=5pt}] (-0.1,-0.4) -- (-2.1,-0.4);
		\draw [decorate,decoration = {brace,amplitude=5pt}] (3.0,-0.4) -- (1,-0.4);
		
		\node at (1.1,-1.0) {\sf\small QFT$_3$ $=$ gravity$_{4}$ for 2 subsectors $\rightarrow$ ($\rm AdS_4$ gravity$)_1$ coupled to ($\rm AdS_4$ gravity)$_2$};
	\end{tikzpicture}
\end{equation}
In the top-down realization \cite{Uhlemann:2021nhu}, the full holographic dual on top now is an $\rm AdS_4$ solution with an internal space which may be thought of as dumbbell shaped. Geometrizing the subsectors separately leads, as indicated on the bottom, to two $\rm AdS_4$ solutions with internal spaces which arise from the two halves of the dumbbell (but are not literally the two halves). They are coupled to each other at their conformal boundaries. Adding black holes now leads to an intermediate description in which a black hole is coupled to a gravitating bath. 

In this work we combine the two ideas. We use the 3d $T_\rho^\sigma[SU(N)]$ SCFT used for wedge holography in \cite{Uhlemann:2021nhu}, which decomposes into two subsectors, as boundary degrees of freedom in a 4d $\mathcal N=4$ SYM BCFT. This leads to the following picture
\begin{equation}\label{eq:double-hol-BCFT-wedge}
	\begin{tikzpicture}
		\node at (1.3,0) {\sf 4d BCFT \ \ $=$ \ \ 3d SCFT$_1$ \ \ \ $+$ \ \ \ 3d SCFT$_2$ \ \ \ $+$ \ \ \ 4d ambient SCFT};
		\draw [decorate,decoration = {brace,amplitude=6pt}] (-2.1,0.4) --  (7.2,0.4);
		\node at (1.8,1.0) {\sf\small QFT$_d$ $=$ gravity$_{d+1}$ for full 4d BCFT $\rightarrow$ asymptotically locally $\rm AdS_5$ gravity};
		
		\draw [decorate,decoration = {brace,amplitude=5pt}] (-0.1,-0.4) -- (-2.1,-0.4);
		\draw [decorate,decoration = {brace,amplitude=5pt}] (3.0,-0.4) -- (1,-0.4);

		\node at (1.1,-1.0) {\sf\small QFT$_3$ $=$ gravity$_{4}$ for subsectors $\rightarrow$ ($\rm AdS_4$ gravity$)_1$ + ($\rm AdS_4$ gravity)$_2$ + 4d ambient CFT};
	\end{tikzpicture}
\end{equation}
Combining the previous discussions and geometrizing all 3 subsectors individually now leads to one $\rm AdS_4$ gravitating system coupled to a second one, which is in turn coupled to the 4d ambient CFT as bath. One can see this as applying double holography twice, or triple holography.\footnote{One can dualize the entire BCFT, or decompose the BCFT into 4d and (all) 3d d.o.f.\ to dualize only the 3d d.o.f. This is double holography. In a second step one can then decompose the 3d d.o.f.\ into two subsectors and geometrize them separately. Altogether we now have 4 different descriptions, 3 of which are gravitational.}
We can introduce black holes of identical temperature into the gravitating systems, and realize the ambient CFT in an excited state at that same temperature. All three systems exchange radiation and due to the balanced temperatures we obtain a steady state.
To send a signal from the first $\rm AdS_4$ gravitating system into the bath, one has to go through the second $\rm AdS_4$ gravitating system.
From the perspective of the bath we get a ``near black hole" and a ``far black" hole. 
Unlike real-world black hole binaries the black holes do not orbit around each other.
The precise holographic and field theory constructions will be discussed in the main part. These more general setups lead to a richer phase structure in terms of entropy curves compared to previous studies, as we will also discuss.

\medskip
\textbf{Outline:} In sec.~\ref{sec:setups} we introduce the setups we work with, including the supergravity solutions, dual field theories, and the implementation of double holography. In sec.~\ref{sec:EE} we determine the HRT/RT surfaces computing the radiation entropy and discuss the expected entropy curves.
We close with a discussion and outlook in sec.~\ref{sec:discussion}.

\section{Black holes coupled to bath}\label{sec:setups}

In this section we discuss the setups we will work with. In sec.~\ref{sec:sugra-sol} we introduce the supergravity solutions, in sec.~\ref{sec:brane-QFT} we discuss the brane setup and dual field theories, and in sec.~\ref{sec:double-hol} we implement their interpretation using double holography. Black holes will be introduced into the setups in sec.~\ref{sec:bh} and their intermediate picture interpretation is discussed in sec.~\ref{sec:bh-int}.

\subsection{Supergravity solutions}\label{sec:sugra-sol}

The supergravity solutions we work with are special cases of a broad class of solutions constructed in \cite{DHoker:2007zhm,DHoker:2007hhe} and further explored in \cite{Aharony:2011yc,Assel:2011xz}. 
We start with a brief discussion of the general features and then focus on the particular solutions of interest here.

The geometry takes the form of a warped product ${\rm AdS_4}\times S^2\times S^2\times\Sigma$ with a Riemann surface $\Sigma$. The solutions involve non-trivial dilaton as well as 2-form and 4-form potentials.
We will only need the Einstein-frame metric, given by
\begin{align}\label{eq:metric-10d}
	ds^2&=f_4^2 ds^2_{\rm AdS_4}+f_1^2 ds^2_{S_1^2}+f_2^2 ds^2_{S_2^2}+4\rho^2 |dz|^2~, 
\end{align}
where $z$ is a complex coordinate on $\Sigma$. The line elements are $ds^2_{\rm AdS_4}$ for unit-radius AdS$_4$ and $ds^2_{S_{1/2}^2}$ for the $S^2$'s. The warp factors are given
in terms of harmonic functions $h_{1/2}$ on $\Sigma$ by
\begin{align}\label{eq:10d-warp-factors}
	f_4^8&=16\frac{N_1N_2}{W^2}~, & f_1^8&=16h_1^8\frac{N_2 W^2}{N_1^3}~, & f_2^8&=16 h_2^8 \frac{N_1 W^2}{N_2^3}~,
	&
	\rho^8&=\frac{N_1N_2W^2}{h_1^4h_2^4}~,
\end{align}
with
\begin{align}
	W&=\partial\bar\partial (h_1 h_2)~, & N_i &=2h_1 h_2 |\partial h_i|^2 -h_i^2 W~.
\end{align}
The ${\rm AdS}_4$ represents the defect conformal symmetry. The solutions preserve 16 supersymmetries and the two $S^2$'s represent the $SU(2)\times SU(2)$ R-symmetry.
Explicit expressions for the remaining Type IIB supergravity fields can be found in \cite{DHoker:2007zhm,DHoker:2007hhe}.

\begin{figure}
	\centering
	\subfigure[][]{\label{fig:sugra-sol}
	\begin{tikzpicture}[scale=1]
		\shade [right color=3dcolor!100,left color=3dcolor!100] (-0.3,0)  rectangle (0.3,-2);
		
		\shade [ left color=3dcolor! 100, right color=4dcolor! 100] (0.3-0.01,0)  rectangle (2,-2);
		\shade [ right color=3dcolor! 100, left color=3dcolor! 100] (-0.3+0.01,0)  rectangle (-2,-2);
		
		\draw[thick] (-2,0) -- (2,0);
		\draw[thick] (-2,-2) -- (2,-2);
		\draw[dashed] (2,-2) -- +(0,2);
		\draw[thick] (-2,-2) -- +(0,2);
		
		\node at (-0.5,-0.5) {$\Sigma$};
		\node at (2.7,-0.65) {\footnotesize $\rm AdS_5/\ZZ_2$};
		\node at (2.7,-1) {\footnotesize $\times$};
		\node at (2.7,-1.35) {\footnotesize $\rm S^5$};

		\draw[very thick] (-1,-0.08) -- (-1,0.08) node [anchor=south] {\footnotesize $N_5/2$ NS5};
		\draw[very thick] (1,-0.08) -- (1,0.08) node [anchor=south] {\footnotesize $N_5/2$ NS5};
		\draw[thick] (-1,-1.92) -- (-1,-2.08) node [anchor=north] {\footnotesize $N_5/2$ D5};
		\draw[thick] (1,-1.92) -- (1,-2.08) node [anchor=north] {\footnotesize $N_5/2$ D5};
		
		\draw (-1,-1.9) -- (-1,-1.75);
		\draw (1,-1.9) -- (1,-1.75);
		\draw [<->] (-1,-1.825) -- (1,-1.825);
		\node at (0,-1.7) {\tiny $2\delta$};
	\end{tikzpicture}
	}
	\hskip 10mm
	\subfigure[][]{\label{fig:BCFT-branes}
	\begin{tikzpicture}[y={(0cm,1cm)}, x={(0.707cm,0.707cm)}, z={(1cm,0cm)}, xscale=1,yscale=1.1]
		\draw[gray,fill=gray!100] (0,0,-0.5) ellipse (1.8pt and 3pt);
		\draw[gray,fill=gray!100] (0,0,1) ellipse (1.8pt and 4.5pt);
		\draw[gray,fill=gray!100,rotate around={-45:(0,0,2.5)}] (0,0,2.5) ellipse (1.8pt and 5pt);
		\draw[gray,fill=gray!100,rotate around={-45:(0,0,4)}] (0,0,4) ellipse (1.8pt and 3pt);				
		
		\foreach \i in {-0.05,0,0.05}{ \draw[thick] (0,-1,-0.5+\i) -- (0,1,-0.5+\i);}
		\foreach \i in {-0.05,0,0.05}{ \draw[thick] (0,-1,1+\i) -- (0,1,1+\i);}

		\foreach \i in {-0.075,-0.025,0.025,0.075}{ \draw (-1.1,\i,2.5) -- (1.1,\i,2.5);}
		\foreach \i in {-0.075,-0.025,0.025,0.075}{ \draw (-1.1,\i,4) -- (1.1,\i,4);}
		
		\foreach \i in {-0.06,-0.03,0,0.03,0.06}{ \draw (0,1.4*\i,-0.5) -- (0,1.4*\i,1);}
		\foreach \i in {-0.1,-0.075,-0.045,-0.015,0.015,0.045,0.075,0.1}{ \draw (0,1.4*\i,1) -- (0,1.4*\i,2.5+\i);}
		\foreach \i in {-0.06,-0.03,0,0.03,0.06}{ \draw (0,1.4*\i,2.5) -- (0,1.4*\i,4);}
		
		\foreach \i in {-0.03,0,0.03}{ \draw (0,1.4*\i,4) -- (0,1.4*\i,6);}
		
		\node at (0,-1.25,-0.5) {\scriptsize $N_5/2$ NS5};
		\node at (0,-1.25,1) {\scriptsize $N_5/2$ NS5};
		\node at (1.0,0.35,2.5) {\scriptsize  $N_5/2$ D5};
		\node at (1.0,0.35,4) {\scriptsize  $N_5/2$ D5};
		\node at (0.2,0.2,1.75) {{\scriptsize $N_{\rm D3}^1$}};
		\node at (0,0.28,0.25) {{\scriptsize $N_{\rm D3}^0$}};
		\node at (0,0.28,3.5) {{\scriptsize $N_{\rm D3}^2$}};
		
		\node at (0,0.28,5.5) {{\scriptsize $N_{\rm D3}^\infty$}};
	\end{tikzpicture}
	}
	\caption{Left: Illustration of the supergravity solutions (\ref{eq:h12}).
		Right: Brane configuration with D3-branes ending on two groups of D5-branes and two groups of NS5-branes. The net numbers of D3-branes ending on each 5-brane differ between the groups for $\delta\neq 0$.\label{fig:sugra-branes}}
\end{figure}
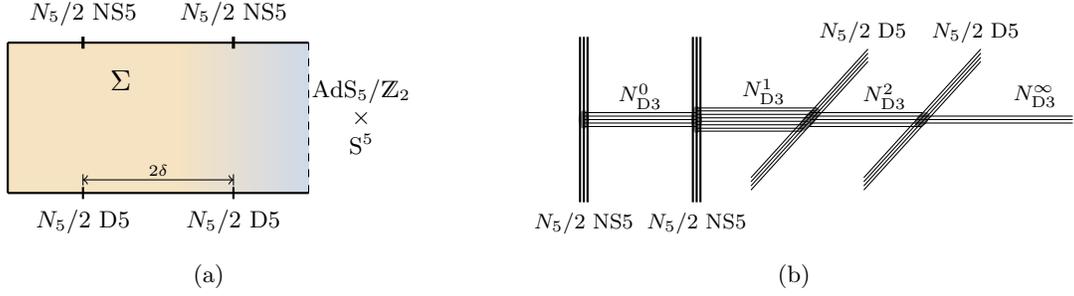

Concrete solutions are specified by a pair of harmonic functions $h_{1/2}$ on $\Sigma$. For the solutions of interest here, $\Sigma$ can be taken as a strip with complex coordinate $z=x+iy$,
\begin{align}\label{eq:Sigma}
	\Sigma&=\left\lbrace z=x+iy\in\CC \ \big\vert \ 0\leq y\leq \frac{\pi}{2}\right\rbrace~.
\end{align}
The relation between the choice of $h_{1/2}$, the brane charges and the dual field theories were worked out for general classes of brane setups in \cite{Aharony:2011yc,Assel:2011xz}. Here we pick a concrete brane setup and BCFT, with associated supergravity solution specified by
\begin{align}\label{eq:h12}
	h_1&=\frac{\pi\alpha'}{4}K e^z-\frac{\alpha'}{4}\frac{N_5}{2}\ln\left[\tanh \left(\frac{z-\delta}{2}\right)\tanh \left(\frac{z+\delta}{2}\right)\right]+\mathrm{c.c.}
	\nonumber\\
	h_2&=-\frac{i\pi\alpha'}{4}Ke^z-\frac{\alpha'}{4}\frac{N_5}{2}\ln\left[\tanh \left(\frac{i\pi}{4}-\frac{z-\delta}{2}\right)\tanh \left(\frac{i\pi}{4}-\frac{z+\delta}{2}\right)\right]+\mathrm{c.c.}
\end{align}
Towards $\Re(z)\rightarrow \infty$ the functions $h_{1/2}$ grow unboundedly, and an asymptotically locally ${\rm AdS_5}\times S^5$ region emerges 
(the $\Re(z)$ direction combines with $\rm AdS_4$ to asymptote to $\rm AdS_5$, while the $\Im(z)$ direction combines with the two $S^2$'s to form an $S^5$ at $\Re(z)\rightarrow \infty$).
This asymptotic ${\rm AdS_5}/\ZZ_2\times S^5$ region provides an $\rm AdS_4$ space as conformal boundary, which is conformally equivalent to a 4d half space and hosts the ambient 4d $\mathcal N=4$ SYM theory.
One of the $S^2$'s collapses on each remaining boundary of $\Sigma$, to form a closed internal space.\footnote{A recent discussion implementing the idea of collapsing cycles in the internal space to realize end-of-the-world branes can be found in \cite{Sugimoto:2023oul}. An attempt leading to unstable solutions can be found in \cite{Harvey:2023pdv}.}
The differentials $\partial h_{1/2}$ have poles on the lower/upper boundary of $\Sigma$ at $\Re(z)=\pm\delta$. These poles correspond to D5/NS5 branes. The locations and residues encode the 5-brane charges and how the D3-branes end on or intersect the 5-branes. 
The solutions are illustrated in fig.~\ref{fig:sugra-sol}.

The solutions in (\ref{eq:h12}) have a single $\rm AdS_4$ as conformal boundary, on which the field theory resides. This $\rm AdS_4$ is conformally equivalent to a half space, and the dual field theory is a BCFT. In terms of computations this is the top-level description in (\ref{eq:double-hol-BCFT}) or (\ref{eq:double-hol-BCFT-wedge}), and the setting where we perform our calculations. Implementing the notion of double holography, i.e.\ the intermediate description, needs more information, which we discuss now.

\subsection{Brane setup and field theory}\label{sec:brane-QFT}

To identify the brane setup described by the solutions (\ref{eq:h12}) and eventually the dual field theory, we need the brane charges. 
We have two groups of $N_5/2$ D5-branes each and two groups of $N_5/2$ NS5-branes each. The number of semi-infinite D3-branes can be obtained by integrating $F_5$ over the 5-cycle formed by a curve connecting $y=0$ to $y=\frac{\pi}{2}$ to the right of all 5-brane sources together with the two $S^2$'s, or by simply reading off the asymptotic $\rm AdS_5\times S^5$ radius. We find
\begin{align}
	N_{\rm D3}^\infty&=2 K N_5 \cosh\delta~.
\end{align}
This leads to the general form of the brane setup shown in fig.~\ref{fig:BCFT-branes}. 
The separation between the 5-branes introduces dimensionful parameters and is only for illustration; the BCFT emerges in the limit where the D5 and NS5-branes all intersect at a point.
What is left to be determined are the numbers of suspended D3-branes, $N_{\rm D3}^{i}$. 

The D3-brane charge going into each 5-brane pole can be determined following  \cite{DHoker:2007zhm,DHoker:2007hhe,Aharony:2011yc,Assel:2011xz}. 
From \cite[(4.28)]{Assel:2011xz} we find
\begin{align}
	Q_{\rm D3}^{(1)}&=\frac{(2\pi)^4}{4} N_5 \left(2 Ke^{-\delta }- (2-\Delta)N_5\right),
	&
	Q_{\rm D3}^{(2)}&=
	\frac{(2\pi)^4}{4} N_5 \left(2  Ke^{+\delta }-\Delta N_5\right),
	\nonumber\\
	\hat Q_{\rm D3}^{(1)}&=\frac{(2\pi)^4}{4} N_5 \left(2 K e^{+\delta }+(2-\Delta) N_5\right),
	&
	\hat Q_{\rm D3}^{(2)}&=\frac{(2\pi)^4}{4} N_5 \left(2  Ke^{-\delta }+\Delta N_5\right),
\end{align}
which correspond, respectively, to the 5-brane sources at $\delta_1=-\delta_2=-\delta$, $\hat\delta_1=-\hat\delta_2=\delta$ in the notation of \cite{Assel:2011xz} (note the ordering).
We defined 
\begin{align}
	\Delta=\frac{1}{2}+\frac{2 }{\pi}\tan ^{-1}\big(e^{-2\delta}\big)~,
\end{align}
which satisfies $\frac{1}{2}\leq\Delta\leq1$ for $\delta\geq0$.
We further have
$Q_{\rm D3}^{(1)} + 	Q_{\rm D3}^{(2)}+\hat Q_{\rm D3}^{(1)}+\hat Q_{\rm D3}^{(2)}=(2\pi)^4N_{\rm D3}^\infty$.
The numbers of suspended D3-branes in fig.~\ref{fig:sugra-branes} are then
\begin{align}
	N_{\rm D3}^0&=\frac{\hat Q_{\rm D3}^{(2)}}{(2\pi)^4}~,
	&
	N_{\rm D3}^1&=\frac{\hat Q_{\rm D3}^{(1)}+\hat Q_{\rm D3}^{(2)}}{(2\pi)^4}~,
	&
	N_{\rm D3}^2&=\frac{Q_{\rm D3}^{(2)}+\hat Q_{\rm D3}^{(1)}+\hat Q_{\rm D3}^{(2)}}{(2\pi)^4}~.
\end{align}
Generally $Q_{\rm D3}^{(2)} \geq Q_{\rm D3}^{(1)}$ and $\hat Q_{\rm D3}^{(1)} \geq \hat Q_{\rm D3}^{(2)}$.
We note that $\hat Q_{\rm D3}^{(1)}$ and $\hat Q_{\rm D3}^{(2)}$ are both positive, while the signs of $Q_{\rm D3}^{(1)}$ and $Q_{\rm D3}^{(2)}$ depend on $K/N_5$ and $\delta$. For $\delta=0$ we have $\Delta=1$ and the sign transitions both happen at $N_5=2K$; this transition was discussed in detail in \cite{Karch:2022rvr}.
For non-zero $\delta$ the sign transitions happen at different values of $N_5/K$.

The form of the brane setup in fig.~\ref{fig:sugra-branes} is natural for the identification of the associated supergravity solution. The field theory description can be made manifest by separating all NS5-branes and using Hanany-Witten transitions to move the D5-branes to a location where they have no net D3-branes attached.
The setup then describes a mixed 3d/4d quiver gauge theory. This is not a conformal theory, as the 3d gauge couplings are dimensionful parameters; the BCFT emerges in the IR limit which drives the 3d gauge nodes to infinite coupling.

We first spell out the quiver for the case where $Q_{\rm D3}^{(1)}$ and $Q_{\rm D3}^{(2)}$ are both negative, so that $N_{\rm D3}^1>N_{\rm D3}^2>N_{\rm D3}^\infty$. This means small enough $K/N_5$. 
We then find
\begin{align}\label{eq:BCFT-quiver-1}
	(s)-\ldots - (&t\cdot s) - (t\cdot s-p)-\ldots - (k)-
 (k+q)-\ldots-(h)-\ldots -(N_{\rm D3}^\infty+r)-\widehat{(N_{\rm D3}^\infty)}
 \nonumber\\
 &\ \ \vert\hskip 75mm \vert
 \nonumber\\
 &\hskip -2mm [N_5/2]
 \hskip 66mm [N_5/2]
\end{align}
The round brackets denote a total of $N_5-1$ 3d unitary gauge nodes. Square brackets stand for 3d flavor symmetries. The hatted node denotes the 4d $\mathcal N=4$ SYM theory on a half space.
The ranks increase in steps of $s=2N_{\rm D3}^0/N_5$ along the first ellipsis, decrease in steps of $p=N_5/2-s$ along the second ellipsis, increase in steps of $q=N_5/2-r$ along the third ellipsis and decrease in steps of $r=(N_{\rm D3}^0-N_{\rm D3}^\infty)/(N_5/2)-N_5$ along the fourth ellipsis.
The remaining parameters are $t=N_5-(N_{\rm D3}^1-N_{\rm D3}^2)/(N_5/2)$, $h=N_{\rm D3}^\infty+r(N_{\rm D3}^2-N_{\rm D3}^\infty)/(N_5/2)$ and $k$ can be determined from the total number of gauge nodes and the above parameters.

Upon changing $N_5/K$, transitions in the shape of the quiver happen at $Q_{\rm D3}^{(1)}=0$ and $Q_{\rm D3}^{(2)}=0$. When both are positive, so that $N_{\rm D3}^1<N_{\rm D3}^2<N_{\rm D3}^\infty$, the quiver takes the form
\begin{align}
	(s)-(2s)-\ldots - (t\cdot s) - (t\cdot s+q)-(t\cdot s+2q)-\ldots - (t\cdot s+(t-1)\cdot q)-\widehat{(N_{\rm D3}^\infty)}
\end{align}
with $s=2N_{\rm D3}^0/N_5$, $t=N_5/2$ and $q=2N_{\rm D3}^1/N_5$. There are no 3d flavors and the rank of the 4d gauge node is not the natural continuation of the ranks of the 3d gauge nodes. Instead of providing flavors for the 3d part of the quiver, the D5-branes now impose boundary conditions on the 4d fields, which break part of the 4d gauge symmetry. When one of the $Q_{\rm D3}^{(a)}$ is negative and one positive, there is one 3d gauge node with flavors and a smaller part of the 4d gauge symmetry is broken by D5-brane boundary conditions. 
For details on the general class of field theories we refer to \cite{Gaiotto:2008sa,Gaiotto:2008sd}.
We focus on the BCFT emerging from (\ref{eq:BCFT-quiver-1}) in the following.

\subsection{Double and triple holography}\label{sec:double-hol}

With the field theories and brane constructions in place, we can implement the notion of double holography following \cite{Karch:2022rvr}. The 
starting point is the mixed 3d/4d quiver gauge theory in (\ref{eq:BCFT-quiver-1}) which flows to a BCFT in the IR. 
We may decompose this quiver as follows,
\begin{equation}\label{eq:quiver-decomp}
\begin{tikzpicture}
	\node [draw, ellipse,align=center] at (0.4,0) {3d SCFT$_1$};
	\node [draw,rectangle,align=center] at (2.4,0) {$k$};
	\draw (1.4+0.4,0) -- (1.75+0.4,0);
	\node [draw,rectangle,align=center] at (0.4,-1.2) {$N_5/2$};
	\draw (0.4,-0.4) -- (0.4,-0.88);
	
	\node [draw, ellipse,align=center] at (6,0) {3d SCFT$_2$};
	\draw (4.25,0) -- (4.6,0);
	\draw (7.4,0) -- (7.8,0);
	\node [draw,rectangle,align=center] at (4,0) {$k$};
	\node [draw,rectangle,align=center] at (8.3,0) {$N_{\rm D3}^\infty$};
	\node [draw,rectangle,align=center] at (6,-1.2) {$N_5/2$};
	\draw (6,-0.4) -- (6,-0.88);

	\node at (11,0.25) {4d $\mathcal N=4$ SYM};
	\node at (11,-0.25) {$SU(N_{\rm D3}^\infty)$};
	
	\draw [decorate,decoration = {brace,amplitude=5pt}] (2.4,0.7) -- (4,0.7);	
	\node at (3.2,1.2) {\small 3d diagonal gauging};
	
	\draw [decorate,decoration = {brace,amplitude=5pt}] (8.1,0.7) -- (11,0.7);	
	\node at (9.5,1.2) {\small gauging via 4d boundary values};
	
\end{tikzpicture}
\end{equation}
The idea is to build up the mixed 3d/4d quiver in (\ref{eq:BCFT-quiver-1}) out of two 3d theories and the 4d $\mathcal N=4$ SYM degrees of freedom on a half space as building blocks.
To define the building blocks we cut the quiver in (\ref{eq:BCFT-quiver-1}) at the $(k)$ 3d gauge node, and at the right end separate out the 4d $SU(N_{\rm D3}^\infty)$ gauge node.
The 3d $(k)$ gauge node is removed, leaving behind the bifundamentals to each side, which remain as fundamental degrees of freedom with respect to the gauge nodes to the left/right of the $(k)$ node; they now carry global flavor symmetries. Likewise, at the right end of the quiver we eliminate the 4d gauge fields, leaving behind fundamental fields attached to the last 3d gauge node. They now carry a global symmetry.
The 3d SCFT sectors in (\ref{eq:quiver-decomp}) then arise as IR limits of genuine 3d quiver gauge theories. These genuine 3d gauge theories are
\begin{subequations}\label{eq:3dquivers}
	\begin{align}\label{eq:3dquiver-1}
		{\rm 3d\ SCFT_1:}\hskip 10mm	(s)-\ldots - (&t\cdot s) - (t\cdot s-p)-\ldots - [k]
		\hskip 12.5mm
		\nonumber\\
		&\ \ \vert
		\nonumber\\
		&\hskip -2mm [N_5/2]
	\end{align}
	and 
	\begin{align}\label{eq:3dquiver-2}
		{\rm 3d\ SCFT_2:}\hskip 10mm
		[k]-(k+q)-\ldots-&(h)-\ldots (N_{\rm D3}^\infty+r)-[N_{\rm D3}^\infty]
		\nonumber\\
		&\ \, \vert
		\nonumber\\
		&\hskip -2mm [N_5/2]
	\end{align}
\end{subequations}
Getting from the three separate sectors in (\ref{eq:quiver-decomp}), (\ref{eq:3dquivers}) back to the BCFT in (\ref{eq:BCFT-quiver-1}) involves two steps: The first is to  diagonally gauge the $U(k)$ flavor groups of the two 3d SCFTs using 3d $U(k)$ gauge fields. The second is to use the boundary values of the 4d $\mathcal N=4$ SYM fields on a half space to gauge the $SU(N_{\rm D3}^\infty)$ global symmetry of the 3d SCFT$_2$.

\begin{figure}
	\centering
		\begin{tikzpicture}[scale=0.9]
		\shade [right color=3dcolor!100,left color=3dcolor!100] (-0.3,0)  rectangle (0.3,-2);
		
		\shade [ left color=3dcolor! 100, right color=3dcolor! 100] (0.3-0.01,0)  rectangle (2,-2);
		\shade [ right color=3dcolor! 100, left color=3dcolor! 100] (-0.3+0.01,0)  rectangle (-2,-2);
		
		\draw[thick] (-2,0) -- (2,0);
		\draw[thick] (-2,-2) -- (2,-2);
		\draw[thick] (2,-2) -- +(0,2);
		\draw[thick] (-2,-2) -- +(0,2);
		
		\node at (-0.5,-0.5) {$\Sigma_1$};

		\draw[very thick] (0,-0.08) -- (0,0.08) node [anchor=south] {\footnotesize $N_5/2$ NS5};
		\draw[thick] (-0.9,-1.92) -- (-0.9,-2.08) node [anchor=north] {\footnotesize $N_5/2$ D5};
		\draw[thick] (0.9,-1.92) -- (0.9,-2.08) node [anchor=north] {\footnotesize $k$ D5};

		\begin{scope}[xshift=65mm]
			\shade [right color=3dcolor!100,left color=3dcolor!100] (-0.3,0)  rectangle (0.3,-2);
			
			\shade [ left color=3dcolor! 100, right color=3dcolor! 100] (0.3-0.01,0)  rectangle (2,-2);
			\shade [ right color=3dcolor! 100, left color=3dcolor! 100] (-0.3+0.01,0)  rectangle (-2,-2);
			
			\draw[thick] (-2,0) -- (2,0);
			\draw[thick] (-2,-2) -- (2,-2);
			\draw[thick] (2,-2) -- +(0,2);
			\draw[thick] (-2,-2) -- +(0,2);
			
			\node at (-0.5,-0.5) {$\Sigma_2$};

			\draw[very thick] (0,-0.08) -- (0,0.08) node [anchor=south] {\footnotesize $N_5/2$ NS5};
			\draw[thick] (-1.4,-1.92) -- (-1.4,-2.08) node [anchor=north,xshift=-2mm] {\footnotesize $k$ D5};
			\draw[thick] (1.5,-1.92) -- (1.5,-2.08) node [anchor=north,xshift=3mm] {\footnotesize $N_{\rm D3}^\infty$ D5};
			\draw[thick] (0,-1.92) -- (0,-2.08) node [anchor=north] {\footnotesize $N_5/2$ D5};
		\end{scope}
	
		\node at (3.25,-1) {$+$};
		\node at (9.3,-1) {$+$};
		\node at (11.5,-1) {4d $\mathcal N=4$ SYM};
	
	\end{tikzpicture}
	\caption{Intermediate description resulting from the decomposition in (\ref{eq:quiver-decomp}) with AdS/CFT applied as in (\ref{eq:double-hol-BCFT-wedge}). The two supergravity solutions are defined by the quivers in (\ref{eq:3dquivers}). They are genuine $\rm AdS_4$ solutions with no asymptotic $\rm AdS_5\times S^5$ regions. The two quivers are different and so are the brane charges, resulting in two different solutions. Both solutions involve D5 and NS5 charges.
		\label{fig:triple-hol-intermediate}}
\end{figure}
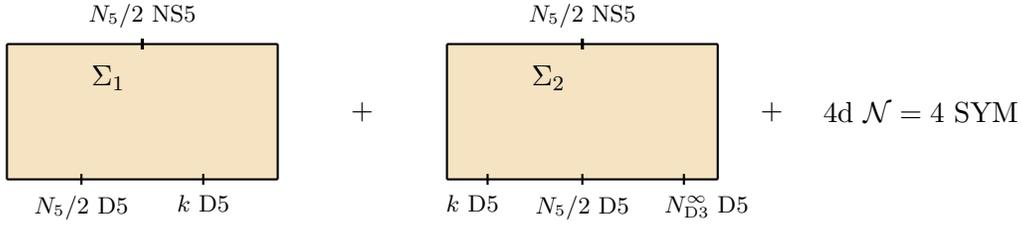

To realize the intermediate picture in the lower line of (\ref{eq:double-hol-BCFT-wedge}) we have to dualize the two 3d sectors in (\ref{eq:quiver-decomp}) separately.
The two 3d SCFTs emerging from the gauge theories in (\ref{eq:3dquivers}) can indeed be dualized into Type IIB $\rm AdS_4$ solutions individually, again using the general $\rm AdS_4\times S^2\times S^2\times\Sigma$ solutions of \cite{DHoker:2007zhm,DHoker:2007hhe}.
The harmonic functions $h_{1/2}$ have to be chosen so as to represent the Hanany-Witten brane setups corresponding to the quivers (\ref{eq:3dquivers}).
The general form for linear 3d quiver SCFTs was discussed in \cite{Assel:2011xz}.
The quivers in (\ref{eq:3dquivers}) are balanced and the general supergravity solutions for this case were spelled out in \cite[sec.~5A]{Coccia:2020wtk}.
For the duals of each 3d theory in (\ref{eq:3dquivers}) we have $N_5/2$ NS5-branes for $N/5-1$ gauge nodes. 
This translates for both to
\begin{align}
	h_2&=-\frac{\alpha'}{4}\frac{N_5}{2}\ln\tanh \left(\frac{i\pi}{4}-\frac{z}{2}\right)+\mathrm{c.c.}
\end{align}
For 3d SCFT$_1$ we have two groups of D5-branes representing the two groups of flavors, leading to
\begin{align}
	h_1&=-\frac{\alpha'}{4}\left[\frac{N_5}{2}\ln\tanh \left(\frac{z-\delta_1}{2}\right)+k\ln\tanh \left(\frac{z-\delta_2}{2}\right)\right]+\mathrm{c.c.}
\end{align}
For 3d SCFT$_2$ we have three groups of D5-branes, leading to
\begin{align}
	h_1&=-\frac{\alpha'}{4}\left[k\ln\tanh \left(\frac{z-\delta_1}{2}\right)+\frac{N_5}{2}\ln\tanh \left(\frac{z-\delta_2}{2}\right)+N_{\rm D3}^\infty\ln\tanh \left(\frac{z-\delta_3}{2}\right)\right]+\mathrm{c.c.}
\end{align}
The solutions are illustrated in fig.~\ref{fig:triple-hol-intermediate}.
The numbers of D3-branes suspended between the 5-branes are determined by the relative positions of the 5-brane sources on $\Sigma$.  
This is captured by the parameters $\delta_i=\ln\tan(\pi {\sf t}_i/N_5)$ where ${\sf t}_i$ denotes the position of the gauge node to which the flavors are attached in the quivers (\ref{eq:3dquivers}) \cite[sec.~5A]{Coccia:2020wtk}.
The 3d duals have no semi-infinite D3-branes and the internal space closes off smoothly at both ends of $\Sigma$, resulting in genuine $\rm AdS_4$ solutions with no asymptotic $\rm AdS_5\times S^5$ regions.
The harmonic functions are generalizations of those for the 3d dual in the $\delta=0$ case in \cite{Karch:2022rvr}.

Fig.~\ref{fig:triple-hol-intermediate} provides the intermediate holographic description for the full BCFT.
Two sectors are gravitational, given by the Type IIB string theory duals of 3d SCFT$_1$ and 3d SCFT$_2$ in (\ref{eq:quiver-decomp}), (\ref{eq:3dquivers}). 
They are of the form $\rm AdS_{4}\times S^2\times S^2\times \Sigma$ with harmonic functions $h_{1/2}$ corresponding to the 3d SCFTs in (\ref{eq:3dquivers}). The two 3d SCFTs differ and so do the holographic duals, which are realized by different harmonic functions $h_{1/2}$. This manifest in the different brane sources and also indicated by the subscripts $\Sigma_{1/2}$ in fig.~\ref{fig:triple-hol-intermediate}.
The third sector is non-gravitational, given by 4d $\mathcal N=4$ SYM, and serves as bath.

The coupling between the 3 sectors can be specified in path integral language following \cite[sec.~3]{Karch:2022rvr}. 
In the decomposition (\ref{eq:quiver-decomp}), the 3d SCFT$_1$ has a global $SU(k)$ symmetry with an associated conserved current, which in turn couples to a 3d background gauge field. We write its partition function as $Z_{\rm 3d\, SCFT_1}[A]$ to highlight the dependence on this background gauge field.
For 3d SCFT$_2$ we similarly highlight the background gauge fields associated with the $SU(k)$ and $SU(N_{\rm D3}^\infty)$ flavor symmetries, as $Z_{\rm 3d\, SCFT_2}[A_1,A_2]$. The full BCFT partition function for  (\ref{eq:quiver-decomp}) can then be written as 
\begin{align}
	Z_{\rm BCFT}&=\int \mathcal DA_{\rm 3d} \mathcal DA_{\rm 4d}\ Z_{\rm 3d\, SCFT_1}[A_{\rm 3d}]\ e^{S_{\rm 3d\, SYM}[A_{\rm 3d}]}\
	Z_{\rm 3d\, SCFT_2}[A_{\rm 3d},A_{\rm 4d}\vert_{\partial}]\
	e^{S_{\rm 4d\,SYM}[A_{\rm 4d}]}\ .
\end{align}
$A_{\rm 3d}$ denotes a 3d  vector field on the boundary of the half space and $A_{\rm 4d}$ denotes the 4d $\mathcal N=4$ SYM vector field with boundary value denoted by $A_{\rm 4d}\vert_\partial$.
The integral over $A_{\rm 3d}$ weighted by the 3d $\mathcal N=4$ SYM action implements the diagonal gauging which couples the two 3d SCFTs. The integral over $A_{\rm 4d}$ weighted by the 4d $\mathcal N=4$ SYM action and with the boundary value as argument in $Z_{\rm 3d\ SCFT_2}$ implements the gauging of the 3d $SU(N_{\rm D3}^\infty)$ global symmetry of 3d SCFT$_2$ using the 4d $\mathcal N=4$ SYM fields.
For the intermediate picture we now apply AdS/CFT in the standard form $Z_{\rm CFT}[{\rm sources}]=Z_{\rm gravity}[{\rm b.c.}]$ to the two 3d sectors.
This leads to 
\begin{align}\label{eq:intermediate-Z}
	Z_{\rm BCFT}&=\int \mathcal DA_{\rm 3d} \mathcal DA_{\rm 4d}\ Z_{\rm IIB\, \Sigma_1}[A_{\rm 3d}]\ e^{S_{\rm 3d\, SYM}[A_{\rm 3d}]}\
	Z_{\rm IIB\,\Sigma_2}[A_{\rm 3d},A_{\rm 4d}\vert_{\partial}]\
	e^{S_{\rm 4d\,SYM}[A_{\rm 4d}]}\,.
\end{align}
$Z_{\rm IIB\, \Sigma_{1/2}}$ denotes the Type IIB string theory partition functions on the $\rm AdS_4\times S^2\times S^2\times \Sigma_{1/2}$ solutions in fig.~\ref{fig:triple-hol-intermediate}. The arguments now highlight the boundary conditions for $\rm AdS_4$ vector fields which are dual to the conserved currents in the 3d SCFTs. These gauge fields emerge from the D5-brane sources. 
This specifies the coupling between the 3 sectors and defines the intermediate description, i.e.\ the bottom line in (\ref{eq:double-hol-BCFT-wedge}). A schematic illustration is in fig.~\ref{fig:radiation-region}.

The decomposition (\ref{eq:quiver-decomp}) with the intermediate description in fig.~\ref{fig:triple-hol-intermediate} can be seen as triple holography, by performing the decomposition in two stages: 
We can first decompose the BCFT into the full 3d degrees of freedom and a 4d bath. The full 3d SCFT then comprises the two quivers in (\ref{eq:3dquivers}) connected by a 3d $(k)$ gauge node. Its dual can be constructed explicitly along the same lines as above. We can then further decompose the 3d SCFT into two subsectors and dualize them separately as described above. Together with the full BCFT dual of sec.~\ref{sec:sugra-sol} we have three different holographic descriptions for the BCFT.

Two special cases connect to previous work: For $\delta=0$ we obtain the BCFT used in \cite{Uhlemann:2021nhu,Karch:2022rvr} to realize a single gravitational sector coupled to a bath.
The two sets of $N_5/2$ flavors in the quiver (\ref{eq:BCFT-quiver-1}) are then at the same node. Decomposing the quiver into the 3d degrees of freedom (without subdividing them) and the 4d ambient CFT leads to the schematic form
\begin{equation}\label{eq:quiver-decomp-deltazero}
	\begin{tikzpicture}
		\node [draw, ellipse,align=center] at (6,0) {3d SCFT};
		\draw (7.3,0) -- (7.8,0);
		\node [draw,rectangle,align=center] at (8.3,0) {$N_{\rm D3}^\infty$};
		\node [draw,rectangle,align=center] at (6,-1.1) {$N_5$};
		\draw (6,-0.4) -- (6,-0.78);

		\node at (11,0.25) {4d $\mathcal N=4$ SYM};
		\node at (11,-0.25) {$SU(N_{\rm D3}^\infty)$};
		
		\draw [decorate,decoration = {brace,amplitude=5pt}] (8.1,0.6) -- (11,0.6);	
		\node at (9.5,1.1) {\small gauging via 4d boundary values};
	\end{tikzpicture}
\end{equation}
Compared to (\ref{eq:quiver-decomp}), the intermediate picture, obtained by dualizing the entire 3d SCFT, comprises a single gravitational sector coupled to 4d $\mathcal N=4$ SYM as bath.
For $K=0$, on the other hand, there are no semi-infinite D3-branes in fig.~\ref{fig:BCFT-branes} and the BCFT reduces to a genuine 3d SCFT. This setup was used in \cite{Uhlemann:2021nhu} for a top-down realization of the wedge holography proposal of \cite{Akal:2020wfl}. The quiver is (\ref{eq:BCFT-quiver-1}) with an empty 4d node.
The 3d quiver was discussed explicitly in \cite[sec.~4.5]{Coccia:2021lpp}.
The decomposition of the 3d SCFT takes the form
\begin{equation}\label{eq:quiver-3d-decomp}
	\begin{tikzpicture}
		\node [draw, ellipse,align=center] at (0.4,0) {3d SCFT$_1$};
		\node [draw,rectangle,align=center] at (2.4,0) {$k$};
		\draw (1.4+0.4,0) -- (1.75+0.4,0);
		\node [draw,rectangle,align=center] at (0.4,-1.1) {$N_5/2$};
		\draw (0.4,-0.4) -- (0.4,-0.78);
		
		\node [draw, ellipse,align=center] at (6,0) {3d SCFT$_2$};
		\draw (4.25,0) -- (4.6,0);
		\node [draw,rectangle,align=center] at (4,0) {$k$};
		\node [draw,rectangle,align=center] at (6,-1.1) {$N_5/2$};
		\draw (6,-0.4) -- (6,-0.78);
		
		\draw [decorate,decoration = {brace,amplitude=5pt}] (2.4,0.5) -- (4,0.5);	
		\node at (3.2,1.0) {\small 3d diagonal gauging};		
	\end{tikzpicture}
\end{equation}
Holographically dualizing the two sectors separately leads to two coupled gravitating systems as intermediate description.
We will use the $\delta=0$ and $K=0$ cases for reference below.

\subsection{Introducing black holes}\label{sec:bh}

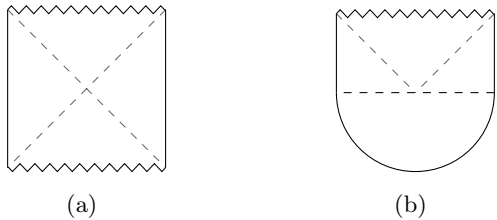
\begin{figure}
	\centering
	\subfigure[][]{\label{fig:bh-prep-1}
	\begin{tikzpicture}[scale=0.7]
		\draw[decorate,decoration = {zigzag,amplitude=1.5pt,segment length=5.65pt}] (0,0) -- (3,0);
		\draw[decorate,decoration = {zigzag,amplitude=1.5pt,segment length=5.65pt}] (3,-3) -- (0,-3);
		\draw (0,0) -- (0,-3);
		\draw (3,0) -- (3,-3);
		\draw[dashed,opacity=0.6] (0,0) -- (3,-3);
		\draw[dashed,opacity=0.6] (3,0) -- (0,-3);
	\end{tikzpicture}
	}\hskip 20mm
	\subfigure[][]{\label{fig:bh-prep-2}
	\begin{tikzpicture}[scale=0.7]
			\draw[decorate,decoration = {zigzag,amplitude=1.5pt,segment length=5.65pt}] (0,0) -- (3,0);
			\draw[dashed] (3,-1.5) -- (0,-1.5);
			\draw (0,0) -- (0,-1.5);
			\draw (3,0) -- (3,-1.5);
			\draw[dashed,opacity=0.6] (0,0) -- (1.5,-1.5);
			\draw[dashed,opacity=0.6] (3,0) -- (1.5,-1.5);
			\draw (0,-1.5) arc (180:360:1.5);
	\end{tikzpicture}
	}
	\caption{Left: Penrose diagram for the two-sided eternal AdS black hole, with the outer edges representing the conformal boundaries. Right: The connection to a Euclidean geometry to prepare a state at $t=0$ via a Euclidean path integral.\label{fig:bh-prep}}
\end{figure}

In this section we introduce black holes into the supergravity solutions of sec.~\ref{sec:sugra-sol}. We will use uncharged, non-rotating $\rm AdS_4$ black branes, which can be introduced into the solutions as follows. So far we have not committed to a choice of $\rm AdS_4$ metric in (\ref{eq:metric-10d}). Though the solutions were derived in \cite{DHoker:2007zhm,DHoker:2007hhe} based on supersymmetry and BPS equations, the only feature needed to show that they solve the Type IIB equations of motion is that $ds^2_{\rm AdS_4}$ describes an Einstein space with negative curvature. We can therefore replace the $\rm AdS_4$ line element in the geometry  (\ref{eq:metric-10d}) throughout with an AdS$_4$ black brane,
\begin{subequations}\label{eq:ads4-bh}
\begin{align}\label{eq:bh-rep}
	ds^2_{\rm AdS_4}& \ \rightarrow \ ds^2_{\rm AdS_4\,bh}=\frac{dr^2}{b(r)}+e^{2r}\left[-b(r)dt^2+ds^2_{\RR^2}\right]~,
\end{align}
with $b(r)=1-e^{3(r_h-r)}$. This geometry describes the exterior of the two-sided AdS$_4$ black hole. The full geometry can be described by the metric
\begin{align}\label{eq:bh-full}
	ds^2_{\rm AdS_4\,bh}&=du^2+e^{2r_h}\cosh^{4/3}\left(\frac{3u}{2}\right)\left[-\tanh^2\left(\frac{3u}{2}\right)dt^2+ds^2_{\RR^2}\right]~.
\end{align}
\end{subequations}
The exterior region covered by the original coordinate $r$ corresponds to $u\in\RR^+$, with the horizon at $u=0$. The extended $\rm AdS_4$ black hole geometry, shown in fig.~\ref{fig:bh-prep-1}, corresponds to a complex contour for $u$, as described e.g.\ in \cite{Hartman:2013qma}.

In the BCFT dual (\ref{eq:h12}) illustrated in fig.~\ref{fig:sugra-sol}, the uniform replacement (\ref{eq:ads4-bh}) leads to a two-sided $\rm AdS_4$ black hole geometry at each point of $\Sigma$. 
This introduces a horizon in the bulk, but also on the asymptotic $\rm AdS_4$ slices at the right end of the strip in fig.~\ref{fig:sugra-sol}, i.e.\ on the conformal boundary where the ambient CFT resides.
In addition to putting the BCFT in an excited state, the replacement thus places the entire BCFT on a fixed, non-dynamical $\rm AdS_4$ black hole geometry at the same temperature as the bulk horizon (we will discuss this from the perspective of the intermediate picture shortly).
The idea then is to prepare the entire system in a pure state at an initial time. The $\rm AdS_4$ black hole geometry has a time reflection symmetry $t\rightarrow -t$, so the geometry can be analytically continued to Euclidean signature. A state at $t=0$ can thus be prepared by a Euclidean path integral, connecting the Lorentzian and Euclidean geometries as shown in fig.~\ref{fig:bh-prep-2}. Time is evolved forward in both exterior regions, which introduces time dependence (the time translation isometry of the Lorentzian eternal black hole would evolve time in opposite directions in the two external regions).

\subsection{Black holes coupled to a bath}\label{sec:bh-int}

\begin{figure}
	\centering
	\begin{tikzpicture}[scale=0.7]
		\draw[decorate,decoration = {zigzag,amplitude=1.5pt,segment length=5.65pt}] (0,0) -- (3,0);
		
		\draw (0,0) -- (0,-1.5);
		\draw (3,0) -- (3,-1.5);
		
		\draw[dashed,opacity=0.6] (0,0) -- (1.5,-1.5) -- (3,0);
		
		\node at (1.5,-3.5) {\small $\rm AdS_{4\,bh}\times S^2\times S^2\times \Sigma_1$};
		\node at (1.5,-4.1) {\small Type IIB string theory};
		
		\node at (-0.25,-0.75) {$a$};
		\node at (3.25,-0.75) {$b$};
		
		\draw[dashed,opacity=0.6] (3,-1.5) -- (0,-1.5);
		\draw (0,-1.5) arc (180:360:1.5);

		\node at (4.5,-1.5) {$+$};
		\node at (10.5,-1.5) {$+$};
		
		\begin{scope}[xshift=60mm]
			\draw[decorate,decoration = {zigzag,amplitude=1.5pt,segment length=5.65pt}] (0,0) -- (3,0);
			
			\draw (0,0) -- (0,-1.5);
			\draw (3,0) -- (3,-1.5);
			\draw[dashed,opacity=0.6] (0,0) -- (1.5,-1.5) -- (3,0);
			
			\node at (1.5,-3.5) {\small $\rm AdS_{4\,bh}\times S^2\times S^2\times \Sigma_2$};
			\node at (1.5,-4.1) {\small Type IIB string theory};

			\node at (-0.25,-0.75) {$c$};
			\node at (3.25,-0.75) {$d$};
			
			\draw[dashed,opacity=0.6] (3,-1.5) -- (0,-1.5);
			\draw (0,-1.5) arc (180:360:1.5);
			
		\end{scope}
		\begin{scope}[xshift=120mm]
			\draw[decorate,decoration = {zigzag,amplitude=1.5pt,segment length=5.65pt}] (0,0) -- (3,0);
			
			\draw (0,0) -- (0,-1.5);
			\draw (3,0) -- (3,-1.5);
			\draw[dashed,opacity=0.6] (0,0) -- (1.5,-1.5) -- (3,0);
			
			\node at (1.5,-3.5) {\small 4d $\mathcal N=4$ SYM};

			\node at (-0.25,-0.75) {$e$};
			\node at (3.25,-0.75) {$f$};
			
			\draw[dashed,opacity=0.6] (3,-1.5) -- (0,-1.5);
			\draw (0,-1.5) arc (180:360:1.5);
			
			\draw [thick,blue] (0.5,-1.5) -- (2.5,-1.5);
			\draw [thick,blue] (0.5,-1.4) -- (0.5,-1.6);
			\draw [thick,blue] (2.5,-1.4) -- (2.5,-1.6);
			\node [blue] at (1.5,-1.9) {$R$};
		\end{scope}
	\end{tikzpicture}
	
	\caption{The decomposition (\ref{eq:quiver-decomp}) leads to the 3 sectors  in fig.~\ref{fig:triple-hol-intermediate}: Type IIB string theory on the $\rm AdS_{4}\times S^2\times S^2\times \Sigma_{1/2}$ geometries dual to 3d SCFT$_{1/2}$ and 4d $\mathcal N=4$ SYM. 
	The replacement (\ref{eq:ads4-bh}) puts all sectors on $\rm AdS_4$ black hole geometries. 	
	The coupling between the duals of 3d SCFT$_1$ and 3d SCFT$_2$ (see (\ref{eq:intermediate-Z})) links the conformal boundaries $a\leftrightarrow c$ and $b\leftrightarrow d$; the coupling between 3d SCFT$_2$ and 4d $\mathcal N=4$ SYM links $c\leftrightarrow e$ and $d\leftrightarrow f$.
	A radiation region $R$ in the field theory bath extending up to a fixed (renormalized) distance from the conformal boundary is shown at the initial time. 
	\label{fig:2bh-bath-Penrose}}
\end{figure}
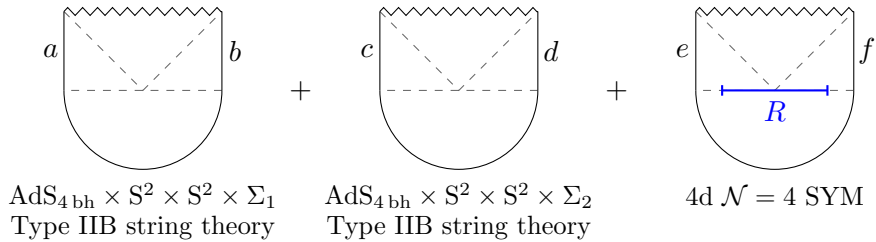

Black holes coupled to a bath emerge when translating  the replacement (\ref{eq:ads4-bh}), which introduces black holes in the full BCFT duals, to the intermediate picture discussed in sec.~\ref{sec:double-hol}.
When $K$ and $\delta$ are both non-zero the intermediate picture comprises 3 sectors, as in (\ref{eq:quiver-decomp}). Applying AdS/CFT as in (\ref{eq:double-hol-BCFT-wedge}) leads two gravitational theories coupled to a bath as in fig.~\ref{fig:triple-hol-intermediate}.
The replacement (\ref{eq:ads4-bh}) in the BCFT dual introduces from the perspective of the intermediate picture a black hole in each of the Type IIB $\rm AdS_4$ duals of 3d SCFT$_1$ and 3d SCFT$_2$ in fig.~\ref{fig:triple-hol-intermediate}. It in addition places the 4d bath CFT on a fixed $\rm AdS_4$ black hole geometry of the same temperature. 
There is one scale in the full BCFT dual which governs all three sectors in the intermediate picture.
The three sectors are coupled, as described in sec.~\ref{sec:double-hol}, and exchange radiation. But with all systems at the same temperature the radiation rates are balanced, leading to a steady state in which none of the black holes evaporate.
Preparing the combined system using a Euclidean path integral leads to the setup in fig.~\ref{fig:2bh-bath-Penrose}.

The (local) bath CFT provides a reservoir with far more degrees of freedom than the number of microstates associated with the combination of black holes in the gravity duals of 3d SCFT$_{1/2}$, which is controlled by the combination of their horizon areas divided by $4G_N$. 
The three systems exchange radiation -- the two gravitational sectors with each other and one of the gravitational sectors with the bath. If the black holes would eternally emit thermal radiation, as suggested by Hawking's calculation, this would build up in the QFT bath and the entropy, e.g.\ of a region $R$ as in fig.~\ref{fig:2bh-bath-Penrose}, would eventually exceed the entropy associated with the black holes in the gravitating sectors.
This is the two-black-hole version of the information paradox for eternal black holes described in \cite{Almheiri:2019yqk}.
Whenever the black hole pair emits some of its Hawking radiation into the QFT bath the question of unitarity arises.\footnote{This is robust regarding the dynamics in the bath. For example, if there is entanglement production in the QFT bath it may contribute to the entropy, but if the entire system is to remain in a pure state the fine-grained entropy would still be bounded by the size of the Hilbert space in the complement of the bath.}

The tension is expected to be resolved by island contributions linked to replica wormhole contributions to the path integral. 
For $\delta=0$ or $K=0$ this was shown in \cite{Uhlemann:2021nhu} (see also \cite{Demulder:2022aij}).
For $\delta=0$, $K>0$ the BCFT takes the form (\ref{eq:quiver-decomp-deltazero}) and the intermediate picture comprises one gravitational system dual to the 3d SCFT coupled to a bath.
The replacement (\ref{eq:ads4-bh}) in the full BCFT dual leads to an intermediate picture as in fig.~\ref{fig:2bh-bath-Penrose} but with the two gravitational sectors merged.
One finds a phase with no islands and growing entropy describing the early phase and a phase with an island in the gravitating sector and constant entropy, which limits the entropy growth and leads to the Page curve for eternal black holes.

For  $K=0$, $\delta>0$ the BCFT reduces to a genuine 3d SCFT which can be decomposed into two sectors as in (\ref{eq:quiver-3d-decomp}). Dualizing them separately leads to two  Type IIB string theory sectors dual to 3d SCFT$_{1/2}$. Introducing black holes via (\ref{eq:ads4-bh}) leads to an intermediate picture with two black holes in gravitational theories which exchange Hawking radiation,
as in fig.~\ref{fig:2bh-bath-Penrose} with the 4d $\mathcal N=4$ SYM bath removed.
Bottom-up versions of this setup were used to model information transfer to a gravitating bath in \cite{Geng:2021mic}.
The system can be split at the conformal boundaries where the two $\rm AdS_4$ geometries are joined, and one can designate one system as black hole and the other as bath.
There is again a phase with no islands and growing entropy as well as a phase with an island and bounded entropy, leading to Page curves \cite{Uhlemann:2021nhu}.

With two gravitating setups hosting black holes coupled in addition to a bath, as in the full fig.~\ref{fig:2bh-bath-Penrose} we can have islands in both of the gravitating systems, in only one of them, or in neither.
This is what we will investigate in sec.~\ref{sec:EE}.
We will refer to the case with an island in only one gravitating sector as partial island.
One would expect steep entropy growth with no islands, slower entropy growth with a partial island, and constant entropy with islands in both gravitating sectors.  The entropy curves could accordingly exhibit more than one transition.

\section{(H)RT surfaces and entropies}\label{sec:EE}

\begin{figure}
	\centering
	\begin{tikzpicture}[scale=0.9]
		\begin{scope}[xshift=-100mm,scale=1.1]
			\draw [lightgray,fill=lightgray] (-0.75,0) ellipse (23pt and 13pt);
			\draw [lightgray,fill=lightgray] (0.75,0) ellipse (23pt and 13pt);
			\node at (-0.75,0) {\small 3d CFT$_1$};
			\node at (0.75,0) {\small 3d CFT$_2$};
			\draw [very thick] (1.55,0) -- (4.3,0);
			\node at (3.0,-0.25) {\small ambient 4d CFT};
			
			\draw [thick,blue] (2.05,0.1) -- (4.3,0.1);
			\draw [thick,blue] (2.05,0) -- (2.05,0.2);
			\node [blue] at (3.2,0.4) {$R$};
		\end{scope}
		
		\draw [very thick] (0,0) -- (2.8,0);
		\node at (1.5,-0.25) {\small ambient 4d CFT};
		
		\draw[fill=lightgray, opacity=0.6,rotate=-10] (-2,0.8) -- (0,0) -- (-2,-0.8);
		
		\draw[fill=lightgray,opacity=0.6,rotate=60] (-2,0.8) -- (0,0) -- (-2,-0.8);
		
		\node at (-1.3,0.2) {$\rm AdS_4^{(1)}$};
		\node at (-0.7,-1.3) {$\rm AdS_4^{(2)}$};
		\draw [thick,blue] (0.5,0.1) -- (2.8,0.1);
		\draw [thick,blue] (0.5,0) -- (0.5,0.2);
		\node [blue] at (1.65,0.4) {$R$};
	\end{tikzpicture}
	\caption{Left: schematic illustration of the BCFT (\ref{eq:BCFT-quiver-1}). Right: Two AdS$_4$ gravity theories coupled at their conformal boundaries to each other and to a non-gravitational 4d CFT. The wedges represent the internal spaces of the $\rm AdS_4\times M^6$ solutions in fig.~\ref{fig:triple-hol-intermediate}. $\rm AdS_4^{(1)}$ is coupled to $\rm AdS_4^{(2)}$ which is in turn coupled to the 4d ambient CFT. The Radiation region $R$ is shown in blue.\label{fig:radiation-region}}
\end{figure}
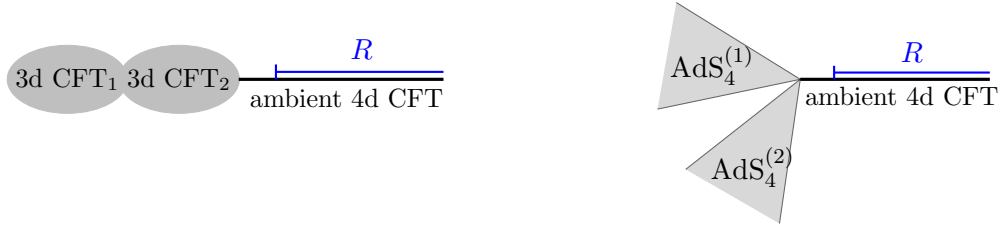

The goal is to compute the entropy in the radiation region shown in fig.~\ref{fig:2bh-bath-Penrose}.
The motivation for considering the entropy in the bath stems from the intermediate picture. But it is also a BCFT question, which can be addressed in any of the three equivalent descriptions in (\ref{eq:double-hol-BCFT-wedge}).
This is shown more schematically in fig.~\ref{fig:radiation-region}.
The virtue of double holography is that we can employ the full BCFT dual in fig.~\ref{fig:sugra-sol} to compute the entropy.
The island rule in the intermediate picture involves an extremization over a QFT entropy and an area term,
\begin{align}
	S_{\rm rad}&=\min_I {\rm ext}_I \left[\frac{{\rm Area}(\partial I)}{4G_N}+S_{\rm semi-cl}\left[\Sigma_{\rm rad}\cup I\right]\right].
\end{align}
The advantage of the full BCFT dual is that both contributions are geometrized. We can compute the radiation entropy without assuming the island rule, which is not proven in dimensions greater than two, but can instead resort to conventional (H)RT surfaces \cite{Ryu:2006bv,Hubeny:2007xt}.
It is their interpretation in the intermediate picture which produces the island rule.

The task then is to compute the entanglement entropy of the region $R$ shown in figs.~\ref{fig:2bh-bath-Penrose} and \ref{fig:radiation-region} in the full BCFT dual shown in fig.~\ref{fig:sugra-sol}.
To this end it is convenient to add the $\rm AdS_4$ radial coordinate to the strip picture in fig.~\ref{fig:sugra-sol}. This leads to the form in fig.~\ref{fig:surfaces.png} below, which shows one exterior region of the $\rm AdS_4$ black hole geometry. The vertical coordinate extends through the horizon and the ER bridge into the second exterior region.
The $\rm AdS_4$ slices at the right end of the strip constitute the conformal boundary (with the $y$-coordinate on the strip representing an angular direction on the $\rm S^5$ in the asymptotic $\rm AdS_5\times S^5$ geometry).
This is the field theory geometry where the RT surfaces are anchored. They extend towards the left end of the strip, which will be discussed in detail below.

To have a dial on how the entropy curve depends on the size of the bath, we take the radiation region as a subset of the 4d $\mathcal N=4$ SYM geometry which extends up to a fixed (renormalized) distance to the interface with the 3d defect (the gravitational sectors in the intermediate picture).
This fixes the vertical anchor point of the surfaces at the right end of the strip in fig.~\ref{fig:surfaces.png}. The picture in the second exterior region is analogous.
For the computations we focus on the central time slice. This gives the starting point of the entropy curve. 
For a single black hole coupled to a bath, the initial RT surfaces are sufficient to determine the qualitative behavior of the Page curve.
With two black holes this gets more interesting.

\subsection{Area functional}

We start with a brief discussion of the area functional and boundary conditions for the surfaces as they approach the boundaries of the strip.
The complete geometries have time reflection symmetry at the initial time slice, and HRT surfaces anchored at this time reduce to RT surfaces in the Euclidean constant-time slice. The computations are performed in the full BCFT dual (\ref{eq:h12}). The ambient CFT geometry emerges at the right end of the strip in fig.~\ref{fig:sugra-sol} at $\Re(z)\rightarrow +\infty$. This is where the RT surfaces are anchored. The choice of radiation region $R$ dictates the anchor as
\begin{align}
	u&=u_R~, & r&=r_R~,
\end{align}
in the $\rm AdS_4$ black hole geometries (\ref{eq:bh-rep}) and (\ref{eq:bh-full}). We will use both coordinate systems.

The surfaces can be described (locally) by the $\rm AdS_4$ radial coordinate $u$ or $r$ as function of the coordinates $(x, y)$ on $\Sigma$ as in (\ref{eq:Sigma}). The area functional for (\ref{eq:bh-rep}) is given by
\begin{align}\label{area_func}
	S&=32\int dx dy \,e^{2r}\left|h_1 h_2 W\right|\sqrt{1+\frac{1}{2b(r)}\left|\frac{h_1 h_2}{W}\right| \left((\partial_x r)^2+(\partial_y r)^2\right)}~.
\end{align}
For a derivation and details see \cite[sec.~3]{Uhlemann:2021nhu}. 
The boundary conditions at points where the surface meets $\partial\Sigma$ are determined from regularity: The 8d surface wraps both $S^2$'s; it can end if the spheres collapse smoothly. This is possible at the boundaries of $\Sigma$, where the $S^2$'s indeed collapse to form a smooth closed 10d geometry. Demanding the 8d surface geometry to also be free from conical singularities leads to Neumann boundary conditions for $r(x,y)$.

The locations on $\Sigma$ where the 5-branes emerge play a special role. The behavior near these sources can be derived analytically and used as a check for the numerics. We again refer to \cite{Uhlemann:2021nhu} for details. Near these points the RT surfaces become sharp, but the contribution to the overall area is negligible. This reflects a general feature in holography for solutions with brane sources: fine-tuned observables will detect the sources and are sensitive to the string theory description there, while sufficiently inclusive observables remain unaffected.\footnote{E.g.\ fundamental Wilson loops can correspond to strings probing the poles \cite{Coccia:2021lpp}. A field theory explanation of the resulting features in matrix models for 5d quiver theories is in \cite[sec.~2.1]{Uhlemann:2020bek}. Robust quantities include free energies and large-rank antisymmetric Wilson loops \cite{Raamsdonk:2020tin,Coccia:2020wtk,Coccia:2021lpp}, as well as the EE's studied here.}

The extremality condition arising from the area functional (\ref{area_func}) is a non-linear PDE with singular points.
It can be discretized and solved using a relaxation method as described in \cite{Uhlemann:2021nhu}. Here we use a different tool, which is the Surface Evolver \cite{Brakke, Brakke1992}. It starts with a trial surface and then iteratively improves it to approximate minimal surfaces using triangulations of increasing resolution. It uses the area functional (\ref{area_func}) locally. 
Some details of the implementation are in app.~\ref{app:surface-evolver}.
As a consistency check we first quantitatively reproduced the results of \cite{Uhlemann:2021nhu} for $\delta=0$, which were obtained using entirely independent methods, and only then applied the Surface Evolver to the  setups with $\delta\neq 0$.

\subsection{(Partial) island and HM surfaces}

In this section we discuss the results of extremizing the area functional (\ref{area_func}). That is, the resulting RT surfaces and their areas. Their interpretation will be discussed in sec.~\ref{sec:surf-interpretation}.

\begin{figure}
	\subfigure[][]{\label{fig:HM}
		\includegraphics[width=0.32\linewidth]{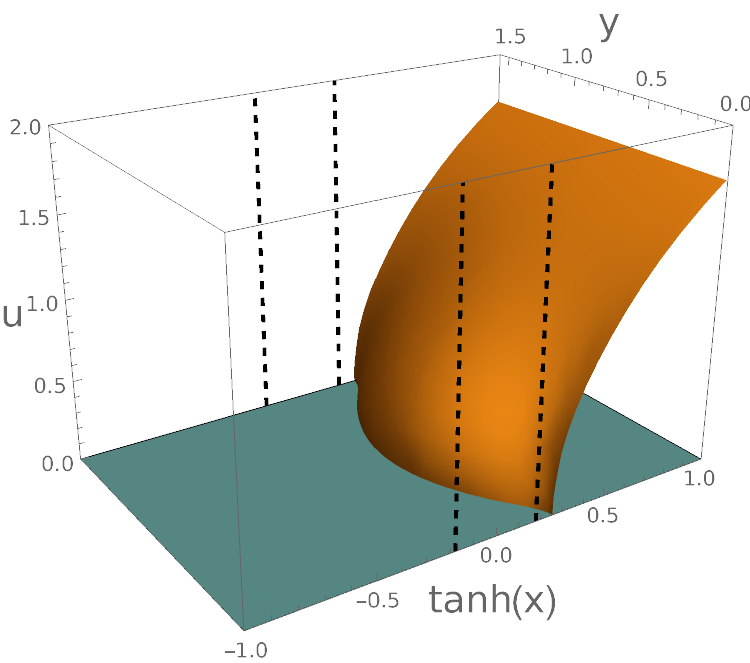}
	}\subfigure[][]{\label{fig:partial-island}
		\includegraphics[width=0.32\linewidth]{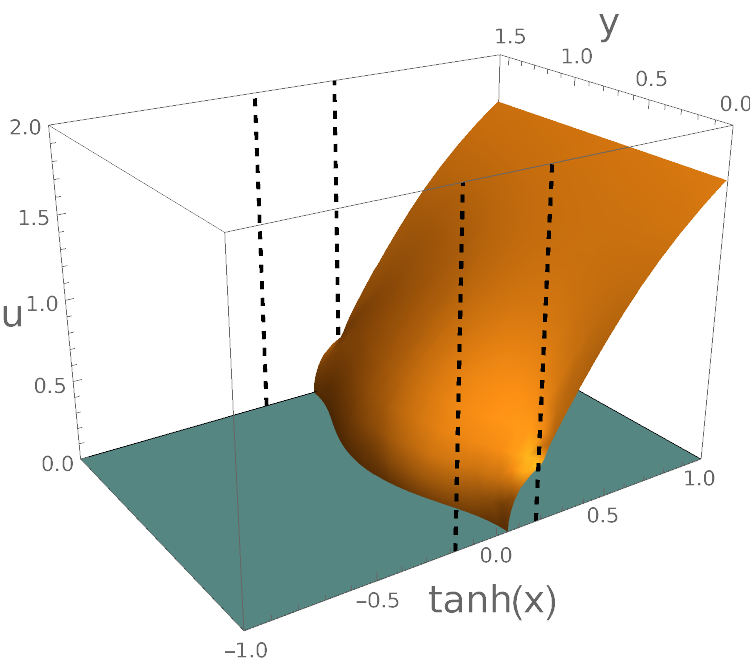}
	}\subfigure[][]{\label{fig:island}
		\includegraphics[width=0.32\linewidth]{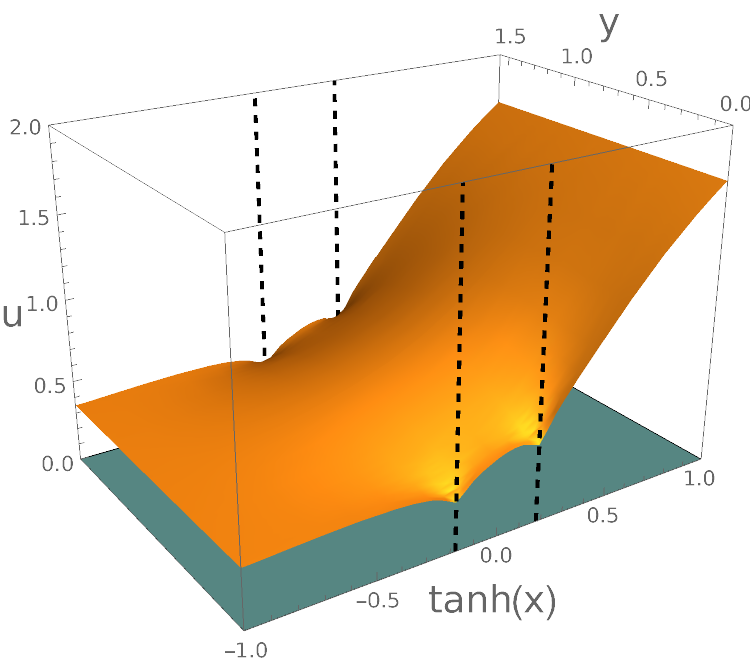}
	}
	\caption{From left to right plots of an HM surface, a partial island surface and a full island surface showing the $\rm AdS_4$ radial coordinate $u$ in the vertical direction and the surfaces as function on the strip $\Sigma$. The horizon is indicated in color as colored plane. The parameters are $\delta=0.18$, $r_R=1.24$ and $N_5/K=1.6$. They are close to the critical point where all three surfaces exist and have comparable area. The vertical dashed posts mark the locations of the poles.}
	\label{fig:surfaces.png}
\end{figure}

Surfaces computing the entropy of the region $R$ in fig.~\ref{fig:2bh-bath-Penrose} are anchored in both exterior regions at the right end of the strip $\Sigma$ in fig.~\ref{fig:surfaces.png} at the same value of the $\rm AdS_4$ radial coordinate outside the horizon. 
There are two options:  The first is that the surface is connected. From the perspective of one exterior region the surface reaches through the horizon into the other exterior region of the black hole geometry and ends on the other boundary of the region $R$.
The second option is that the surface has two components, one in each exterior region, which stay outside the horizon and close off smoothly in the exterior region, at the boundaries of $\Sigma$ where the 2-spheres in the geometry collapse.
These are topologically distinct configurations.

Both options were realized in the discussion of \cite{Uhlemann:2021nhu} for $\delta=0$ when there is only one pair of 5-brane sources. In that case, there are HM surfaces (named in reference to \cite{Hartman:2013qma}) dropping into the horizon before reaching the region of $\Sigma$ with the 5-brane sources.
There are also surfaces reaching past the 5-brane sources. At $\delta=0$, if surfaces reach past the 5-brane sources outside the horizon, they reach all the way to the left end of $\Sigma$ where they close off smoothly.
These are the island surfaces with constant area.

For the setups with $\delta>0$ and two pairs of 5-brane sources, we find an additional option. Surfaces can reach past the first pair of 5-brane sources but drop into the horizon before reaching the second pair. We denote these options, shown in fig.~\ref{fig:surfaces.png}, as follows:
\begin{itemize}
	\item[(i)] The surface drops into the horizon before reaching any 5-brane sources and connects to an identical copy in the second exterior region.
	We refer to this type as HM surface.
	\item[(ii)] The surface extends past the first pair of 5-brane sources but drops into the horizon and connects to its mirror before reaching the second pair of 5-brane sources.
	\item[(iii)] The surface stays outside the horizon entirely and closes off smoothly in the exterior geometry at the boundaries of $\Sigma$. We refer to this type as island surface.
\end{itemize} 
Option (ii) only arises for $\delta>0$. It shares features with both (i) and (iii): it is similar to the HM surface (i) in the sense that it reaches through the horizon, and it shares with the island surface (iii) that it reaches past at least some 5-brane sources. We could call it a ``long" HM surface, but anticipating the discussion in sec.~\ref{sec:surf-interpretation} we will call type (ii) a {\it partial island} surface.

Whether and which of these surfaces exist depends on the parameters $(N_5,K,\delta,u_R)$. 
Examples are shown in fig.~\ref{fig:surfaces.png} for a choice of parameters where all three surfaces exist.
These surfaces are at the initial time slice, when the two exterior regions are connected directly.

\subsection{Phase diagram}\label{sec:phase-diag}

In this section we discuss when the different types of surfaces exist and how their areas compare, to construct a form of phase diagram.

\begin{figure}
	\centering
	\subfigure[][]{\label{fig:short_hm_existence}
		\includegraphics[width=0.42\textwidth]{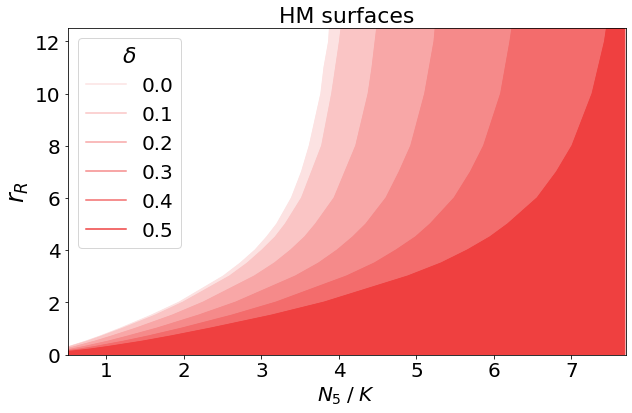}
	}
	\hskip 8mm
	\subfigure[][]{\label{fig:long_hm_existence}
		\includegraphics[width=0.40\textwidth]{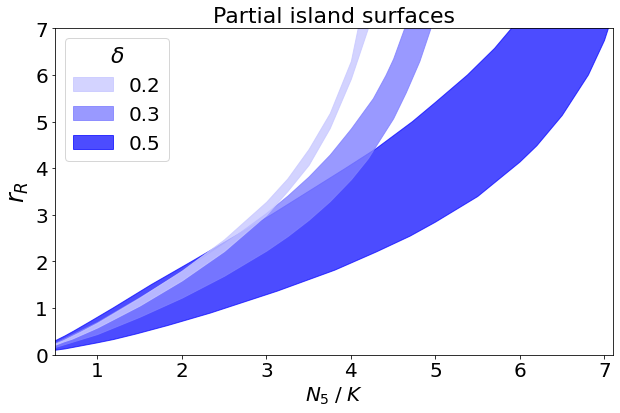}
	}
	\subfigure[][]{	\label{fig:delta=0.3_short_long.png}
		\includegraphics[width=0.41\textwidth]{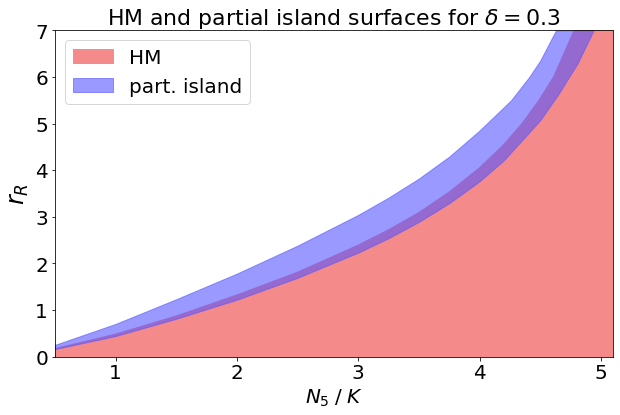}
	}
        \hskip 8mm
        \subfigure[][]{	\label{fig:delta=0.5_short_long.png}
		\includegraphics[width=0.42\textwidth]{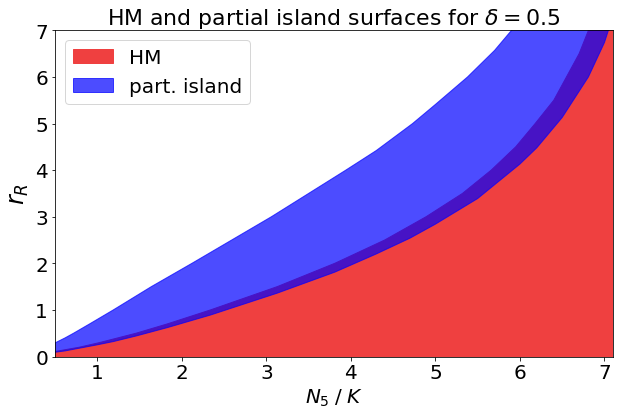}
	}
	\caption{Top left: Regions where HM surfaces are found for various $\delta$ shaded in red. All regions extend towards infinite $N_5/K$; they overlap.  The shape of the regions suggests the existence of a critical $N_5/K$ for each $\delta$ beyond which HM surfaces exist for any value of $r_R$. Top right: Regions where partial island surfaces are found. They take the form of bands in the $(N_5/K,r_R)$ plane. 
	Bottom: Regions where partial island and HM surfaces are found for $\delta=0.3$ (left) and $\delta=0.5$ (right). The regions were included separately in figs.~\ref{fig:short_hm_existence} and \ref{fig:long_hm_existence}, and are now shown together. They overlap in a thin band. The full island surface exists across the entire parameter space.}
\end{figure}

The HM surfaces (fig.~\ref{fig:HM}) and full island surfaces  (fig.~\ref{fig:island}) are qualitatively similar to the $\delta=0$ case. At finite temperature there always is an island surface,\footnote{At zero temperature the existence of the island surface depends on the parameters. But at zero temperature the area of the HM surface is also time-independent, so the entropy is constant either way.} while the existence of the HM surface depends on the parameters. There are two regimes: At small $N_5/K$, the HM surface exists only for sufficiently small $u_R$. For large enough $N_5/K$, the HM surface always exists. This is illustrated in fig.~\ref{fig:short_hm_existence}. For $\delta=0$ the `critical value' separating the two regimes is $N_5/K\approx 4$. This critical value increases with $\delta$.
The partial island surfaces in fig.~\ref{fig:partial-island} only exist in certain bands in parameter space, as shown in fig.~\ref{fig:long_hm_existence}.
As $\delta$ decreases and the poles become closer, $u_R$ and $N_5/K$ must be more finely tuned to find these surfaces, shrinking the width of the band. We did not find partial island surfaces below $\delta\sim 0.13$.

With the partial island and HM surfaces only existing in certain regions of parameter space, one may wonder whether these regions overlap. This is indeed the case. As illustrated in fig.~\ref{fig:delta=0.3_short_long.png} and \ref{fig:delta=0.5_short_long.png} the regions overlap in a narrow band.
The overlap is wider for larger $N_5/K$, and partial islands may be found past the critical $N_5/K$ for HM surfaces. The overlap also grows with increasing $\delta$, although not as much as the total width of the partial island band. The overlap region constitutes a smaller fraction of the full band for larger $\delta.$

The entanglement entropy is determined by the surface with minimal area among all candidate surfaces. So we need to compare the areas.
As mentioned above the full island surface always exists at finite temperature, providing an upper bound on the entropy and preventing unbounded growth. The partial island and HM surfaces, on the other hand, both stretch through the horizon and their area grows with time as a result. Which of these surfaces is dominant at the initial time is a crucial part in determining the shape of the entropy curve.

\begin{figure}
	\centering
	\includegraphics[width=0.48\textwidth]{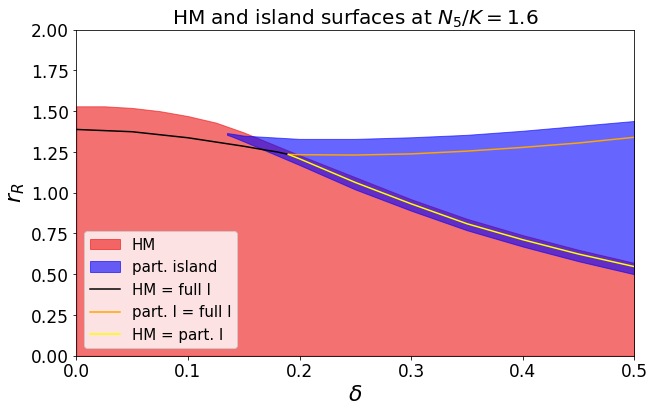}\hskip 5mm
	\includegraphics[width=0.48\textwidth]{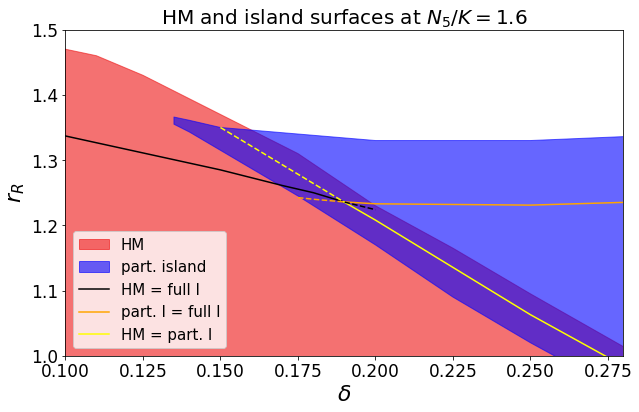}
	\caption{Phase diagram for surfaces at $N_5/K=1.6$. The full island surface exists throughout and is not shown explicitly. The black, orange and yellow curves mark the boundaries of dominance between the three types of surfaces. Partial island surfaces have the smallest area in the wedge-shaped region between the yellow and orange curves, whereas HM and full island surfaces dominate at smaller and larger $r_R$ respectively. Below/above the black curve the HM/island surface is dominant. The boundaries are consistent with the existence of a triple point near $\delta\sim 0.19$ and $r_R\sim 1.235$.}
	\label{fig:phase_diagram.png}
\end{figure}

Fig.~\ref{fig:phase_diagram.png} shows a $(\delta,r_R)$ section of the parameter space at fixed $N_5/K$. The blue region shows how the partial island region shrinks with decreasing $\delta$ and closes off near $\delta\sim 0.135$. HM surfaces exist for sufficiently small $r_R$ throughout the plot ($N_5/K$ is below the critical value set by fig.~\ref{fig:short_hm_existence}).
In the upper white region only the full island exists and is therefore dominant.
Upon decreasing $r_R$ at fixed small $\delta$, one enters a regime where island and HM surfaces coexist. Along the black curve their areas are identical. The island surface stays dominant in the upper part of the coexistence region, but for small $r_R$ the HM surface is dominant. This is qualitatively similar to the $\delta=0$ case.
For large $\delta$, decreasing $r_R$ first leads to a regime where the partial and full island surfaces coexist. Along the orange curve their areas are identical. Above it the island surface is still dominant while below it the partial island surface has smaller area. 
Further decreasing $r_R$ leads to a band of $r_R$ values where all three surfaces coexist. In this band the question of dominance is between the partial island surface and the HM surface. The yellow curve shows the points where their areas are equal; above it the partial island dominates, below it the HM surface is dominant.
A notable feature of the phase diagram is the triple point at $(\delta,r_R) \sim (0.19,1.235)$ at which the area differences between all three surfaces become negligible. At this point the three curves indicating equal areas between the island and HM surface, between the island and partial island surfaces, and between partial island and HM surfaces all meet.
The curves continue past the critical point (though they then compare sub-dominant surfaces, as denoted by the dashed lines.) There is no need for them to all meet at one point, yet at least at the level of our numerics they do.

\subsection{Interpretation}\label{sec:surf-interpretation}

Based on the phase diagram of surfaces discussed sec.~\ref{sec:phase-diag} for the HM, partial island and island surfaces, there are several natural scenarios for the time evolution of the entropy.

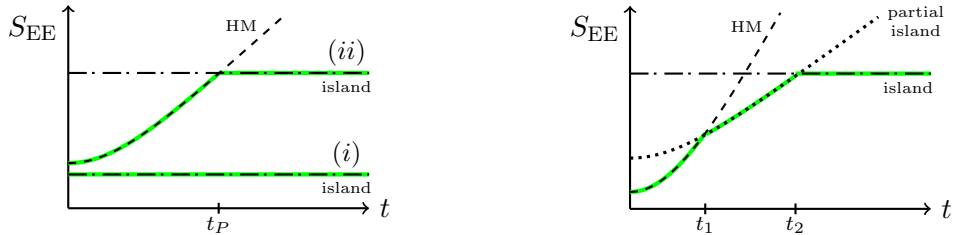
\begin{figure}
	\centering
	\begin{tikzpicture}[yscale=0.9]
		\draw [->,thick] (0,0) -- (4,0) node[anchor=west] {$t$};
		\draw [->,thick] (0,0) -- (0,3) node[anchor=north east] {$S_{\rm EE}$};

		\draw [green,ultra thick] (0,0.5) -- +(4,0);
		\draw [thick, dash pattern={on 7pt off 2pt on 1pt off 3pt}] (0,0.5) -- +(4,0);
		
		\node at (3.7,0.8) {\small $(i)$};
		
		\draw[domain=0:2.0, smooth, variable=\x,ultra thick,green] plot ({\x}, {2/3+ln(cosh(\x))});
		\draw [green,ultra thick] (2.0,2) -- (4,2);
		
		\draw [thick,dash pattern={on 7pt off 2pt on 1pt off 3pt}] (0,2) -- +(4,0);

		\draw[domain=0:2.85, smooth, variable=\x,dashed, thick] plot ({\x}, {2/3+ln(cosh(\x))});
		
		\node at (3.7,2.3) {\small $(ii)$};	
		
		\draw [thick ](2,+0.08) -- (2,-0.08) node [anchor=north,yshift=1mm] {\scriptsize $t_P$};
		
		\node at (2.3,2.7) {\tiny HM};
		\node at (3.7,1.8) {\tiny island};
		\node at (3.7,0.3) {\tiny island};
	\end{tikzpicture}
\hskip 20mm
	\begin{tikzpicture}[yscale=0.9]
		\draw [->,thick] (0,0) -- (4,0) node[anchor=west] {$t$};
		\draw [->,thick] (0,0) -- (0,3) node[anchor=north east] {$S_{\rm EE}$};

		\draw[domain=1.01:2.2, smooth, variable=\x,green,ultra thick] plot ({\x}, {3/4+8/10*ln(cosh(\x))});	
		
		\draw[domain=0:1, smooth, variable=\x,ultra thick,green] plot ({\x}, {1/4+2*ln(cosh(\x))});
		\draw [green,ultra thick] (2.2,2) -- (4,2);

		\draw [thick,dash pattern={on 7pt off 2pt on 1pt off 3pt}] (0,2) -- +(4,0);

		\draw[domain=0:2, smooth, variable=\x,dashed, thick] plot ({\x}, {1/4+2*ln(cosh(\x))});	
		
		\draw[domain=0:3.3, smooth, variable=\x,dotted, very thick] plot ({\x}, {3/4+8/10*ln(cosh(\x))});	
		
		\draw [thick ](1,+0.08) -- (1,-0.08) node [anchor=north,yshift=1mm] {\scriptsize $t_1$};
		
		\draw [thick ](2.2,+0.08) -- (2.2,-0.08) node [anchor=north,yshift=1mm] {\scriptsize $t_2$};
		
		\node at (1.55,2.7) {\tiny HM};
		\node at (3.7,1.8) {\tiny island};
		\node at (3.8,2.9) {\tiny partial};
		\node at (3.8,2.63) {\tiny island};
	\end{tikzpicture}
	\caption{Left: Sketch of entropy curves without partial island surface: $(i)$ flat if the island surface (dot-dashed) dominates from the outset $(ii)$ initial growth before saturating to a finite value consistent with unitarity if the HM surface (dashed) dominates initially. Right: With partial island surface (dotted) we can have 3 phases, an initial no-island phase with steep entropy growth, a partial island phase with slower entropy growth and a final full island phase with constant entropy.
	Depending on the relative starting points of the 3 curves (cf.\ fig.~\ref{fig:phase_diagram.png}) and their slopes there can also be fewer transitions.
	\label{fig:entropy-curve-sketches}}
\end{figure}

The simplest is where the island surface dominates from the outset, e.g.\ in the upper part of fig.~\ref{fig:phase_diagram.png}.
This leads to a constant entropy curve (e.g.\ option (i) on the left in fig.~\ref{fig:entropy-curve-sketches}). The preparation of the initial state in fig.~\ref{fig:2bh-bath-Penrose} leads to entanglement between the two black hole systems and with the bath at the initial time. In the regime where the island surface dominates from the outset, this initial entanglement appears to favor the inclusion of a full island right away. The flat entropy curve is bounded and consistent with unitarity.

A second option is when the HM surface dominates initially and partial island surfaces play no role, e.g.\ at small $\delta$.\footnote{We note that the existence of the partial island and HM surfaces may depend on time. The plot in fig.~\ref{fig:phase_diagram.png} only shows the phase diagram at the initial time. So the discussion of the time dependence is qualitative.} 
The area of the HM surface grows in time, representing an increasing entropy. At late times the growth is linear \cite{Hartman:2013qma}.
When the HM surface outgrows the island surface  the island surface with constant area becomes dominant. This competition leads to the Page curve for eternal black holes (option (ii) on the left in fig.~\ref{fig:entropy-curve-sketches}) consistent with unitarity. 
Since the black holes do not evaporate the entropy curve does not decrease back to the initial value. The paradox prevented by the island surface is that of eternal entropy growth (discussed in sec.~\ref{sec:bh-int}).
An explicit calculation of Page curves for $K=0$ can be found in \cite[fig.~12(a)]{Uhlemann:2021nhu}.
A similar shape for the entropy curve arises if the partial island surface dominates initially, grows over time and then transitions to the island surface.
Though this option represents different entanglement structure with partial island from the outset, the entropy curve would also have a growing regime transitioning directly to saturation.

A qualitatively new scenario arises for large enough $\delta$, when island, HM and partial island surfaces can all play a role.
The partial island surfaces stretch through the ER bridge into the second exterior region of the $\rm AdS_4$ black hole geometry, leading to an area which grows with time. The late-time behavior should be linear based on the arguments of \cite{Hartman:2013qma} (see also \cite[fig.~12(a)]{Uhlemann:2021nhu}),\footnote{At late times we expect the connection through the ER bridge, which then is very long, to be governed by surfaces similar to those studied in \cite{Uhlemann:2023oea}. The time evolution would then follow the discussion in \cite[sec.~6]{Uhlemann:2021nhu}.} but we expect a more gentle slope.
This suggests the option of entropy curves with two transition times: starting out with the HM surface and steep growth, transitioning to a partial island phase with more gradual growth and eventually to a full island phase with constant entropy.\footnote{Entropy curves of qualitatively similar shape appeared in \cite{Balasubramanian:2021xcm}.}  These options are sketched in fig.~\ref{fig:entropy-curve-sketches}.
Which options for the entropy curve are available depends on the relative starting points of the area curves, discussed in fig.~\ref{fig:phase_diagram.png}.

An interesting aspect is the precise interpretation of the partial island surface in the intermediate picture. In the full BCFT dual the partial island surface extends past the first pair of 5-brane sources outside the horizon but not past the second pair (cf.\ fig.~\ref{fig:partial-island}). 
The gravitational sector at the center of the intermediate picture in fig.~\ref{fig:triple-hol-intermediate} is built around the NS5 and D5 branes at the right poles in fig.~\ref{fig:sugra-sol}, where the partial island surface is still outside the horizon, while the gravitational sector on the left in  fig.~\ref{fig:triple-hol-intermediate} is built around the NS5 and D5 branes corresponding to the left poles in fig.~\ref{fig:sugra-sol}, which are not captured by the partial island surface. This follows from the decomposition in (\ref{eq:quiver-decomp}), (\ref{eq:3dquivers}).
This suggests that the partial island surface corresponds to having an island in the gravitational sector in the center in fig.~\ref{fig:triple-hol-intermediate}, but not in the one on the left.\footnote{%
In general the relation between the geometry of the full BCFT dual and the geometries in the intermediate picture is non-trivial, as emphasized in \cite{Karch:2022rvr}. It simplifies in certain limits, which here amount to large $\delta$ and small $K/N_5$. In this limit we expect the discussion to be accurate. For more general parameters the arguments should be seen as	approximate. We may have a large island in one sector and a small island in the other.}
From the perspective of an observer in the bath this would be a phase with an island in the `near' black hole but not in the `far' black hole.\footnote{%
We recall that the notion of near and far emerges from the way the two black holes are coupled to the bath: the radiation from the far black hole has to go through the near black hole to reach the bath.}%

\section{Discussion}\label{sec:discussion}

What potentially observable signatures of quantum gravity can one hope to extract from black hole binaries?
Motivated by this question we constructed setups which, via double holography, realize pairs of black holes coupled to a bath in Type IIB string theory. 
Each black hole lives in its own AdS$_4$ space, and the AdS$_4$ spaces are coupled at their conformal boundaries. The gravitational attraction between the black holes is balanced by the confining nature of AdS.
The setups differ from those observed by gravitational wave experiments but they provide a starting point for studying pairs of black holes. In the spirit of recent black hole discussions we coupled the combined black hole system to a non-gravitational bath where the radiation extracted from the combined system can be studied. As a first step we focused on the fine grained or von Neumann entropy.

In scenarios with a single black hole coupled to a bath the entropy curve emerges from the competition between a no-island phase with growing entropy and an island phase with constant entropy.  For a pair of black holes we may expect more options, with islands in one of the two black hole systems, in none, or in both.
We used double holography to evaluate the radiation entropy, which allows us to use standard Ryu-Takayanagi surfaces without having to assume the island rule.
Working in top-down models further allowed us to implement a precise notion of double holography with a concrete holographic dictionary.
Focusing on the initial-time surfaces, we indeed found a new option of having a partial island.
From the perspective of the bath there is a near black hole and a far black hole, and the partial island is in the near black hole.
We determined the phase diagram of surfaces and their dominance at the initial time in dependence on the relative size and interactions between the black holes and with the bath. This revealed for each type of surface regimes where it is dominant. We also  found a triple point where all surfaces have equal area.
The qualitative shape of the entropy curve depends on the competition between island, partial-island and no-island surfaces at the initial time, and our results suggest that entropy curves can in general have two transitions  rather than a single Page time, going through no-island, partial-island and full-island phases.
It would be interesting to study the full time evolution more quantitatively.
The top-down setups also provide an ideal setting to connect the Page curve discussions to the black hole microstate program. The Page curve discussions here and in \cite{Uhlemann:2021nhu} are based on finite-temperature black holes, but the setups also provide a setting where the entropy of extremal black holes can be explained as statistical entropy in terms of microstates \cite{Coccia:2020cku,Coccia:2020wtk}.

The notion of double holography employed in this work is an elaboration on the implementation in \cite{Karch:2022rvr}. 
In keeping with previous terminology it can be called triple holography. Double holography starts with a BCFT where one can either dualize the entire BCFT into a gravitational description, or isolate the boundary degrees of freedom and dualize them alone. The latter leads to the `intermediate' holographic description. In our setup we can first isolate the 3d boundary degrees of freedom out of a 4d BCFT and dualize them separately, leading to one intermediate description, but we can then further decompose the 3d degrees of freedom into two subsectors which are dualized separately. We end up with 3 holographic descriptions: the full BCFT dual, the full 3d dual coupled to the 4d ambient CFT, and the description with two 3d duals for the two subsectors which are coupled to each other and to the 4d ambient CFT. This last description is the physically interesting one. The full BCFT dual is convenient for computations.
All descriptions can be made precise thanks to the breadth of the D3/D5/NS5 brane setups \cite{Gaiotto:2008sa,Gaiotto:2008sd} and the associated supergravity solutions \cite{DHoker:2007zhm,DHoker:2007hhe,Aharony:2011yc,Assel:2011xz}. 
Our construction is based on the solutions in fig.~\ref{fig:sugra-sol} with two pairs of D5/NS5 branes, which naturally leads to a 3d SCFT comprising two sectors with individual duals supported by D5 and NS5 sources. The construction can be generalized, e.g.\ to multiple pairs of D5/NS5 pairs leading to an intermediate picture with multiple gravitational systems coupled to a bath.
This is a somewhat orthogonal approach to seeking multiple black hole in a single AdS \cite{Monten:2021som,Cai:2022qac}.

It would be interesting to complement the top-down constructions with bottom-up brane\-world realizations, e.g.\ using multiple effective branes separating AdS spaces and wedges with different radii (similar to the constructions in \cite{Bachas:2022etu,Yadav:2023qfg}), to combine the wedge holography of \cite{Akal:2020wfl} with the bottom-up braneworld double holography of \cite{Karch:2000gx,Karch:2000ct}.
Conversely, it would be interesting to study aspects which have  been studied recently in bottom-up models in the top-down models, e.g.\ additional information-theoretic quantities \cite{Hernandez:2020nem,Ling:2021vxe,Kawabata:2021vyo,Li:2021dmf,Basu:2022reu,Afrasiar:2022ebi,Lu:2022cgq,Basu:2022crn,Basu:2023wmv,Aguilar-Gutierrez:2023ccv,Liu:2022pan}, 
charged black holes \cite{Ling:2020laa,Yadav:2022fmo,Xu:2023fad,Jeong:2023lkc},  connections to bootstrap conditions \cite{Geng:2021iyq,Reeves:2021sab,Izumi:2022opi,Kusuki:2022ozk,Geng:2023iqd,Matsuo:2023cmb} and effective brane description \cite{Ghosh:2021axl,Grimaldi:2022suv,Hu:2022ymx,Miao:2022mdx,Perez-Pardavila:2023rdz,Huertas:2023syg,Kanda:2023zse,Jain:2023xta}, more general dynamics \cite{Geng:2022slq,Emparan:2023dxm,Chou:2023adi,Yu:2023whl,Deng:2023pjs}, and connections to de~Sitter space and cosmology \cite{Geng:2021wcq,Hartman:2020khs,Balasubramanian:2020xqf,VanRaamsdonk:2021qgv,Antonini:2022blk,Yadav:2022jib,Goswami:2022ylc,Emparan:2022ijy,Aguilar-Gutierrez:2023tic,Chang:2023gkt,Goswami:2023ovb}.

\section*{Acknowledgments}
This work is supported in part by the U.S. Department of Energy under grant DE-SC0007859. LPZ gratefully acknowledges support from an IBM Einstein fellowship during 2022/2023 while at the Institute for Advanced Study.

\appendix	
	
\section{Surface Evolver}\label{app:surface-evolver}

The Surface Evolver starts with a trial surface and then iteratively improves it to approximate minimal surfaces using triangulations of increasing resolution. The program supports user-defined boundary conditions and non-Euclidean metrics. For technical details on the implementation of area functionals such as \eqref{area_func}, see \cite{Deddo:2022wxj}. We will be brief on the technical description here, but include a detailed quantitative comparison to \cite{Uhlemann:2021nhu} to show that the tool works as intended.

For the numerical computation we work with a rescaled coordinate $X=\tanh{x}$. Taking advantage of the reflection symmetry $y\rightarrow \frac{\pi}{2}-y$, we only compute the $y\leq\frac{\pi}{4}$ half of the surface. Cutoffs are implemented at $X=\pm 0.995$ as well as at $y=0.01$. The latter avoids numerical issues due to the surface forming cusps near the 5-branes. We verified that the effect of varying the $y$ cutoff on the areas is negligible.

All surfaces are anchored at a fixed value of the $\rm AdS_4$ radial coordinate, $u=u_R$, at $X=0.995$, with $X=1$ corresponding to the right end of $\Sigma$. HM surfaces are permitted to drop down to the horizon at $u=0$. The resulting minimal surfaces are consistent with Neumann boundary conditions at the horizon, so we did not impose this as an extra constraint. For both HM and island surfaces, Neumann conditions at the $y$ boundaries are also automatically satisfied. One may verify this behavior by connecting the surface to a duplicate across the boundary and observing that the evolution of the double surface is identical to the original within machine precision.
However, for the island surfaces which reach all the way to $X=-0.995$, the proper boundary conditions must be imposed by hand. This is accomplished approximately by forcing the normal vectors of the last row of triangles to point exactly in the $u$ direction. As the triangle size decreases, this approximation improves.

\begin{figure}
    \centering
    \subfigure[][]{\label{fig:del0comp}
        \begin{tikzpicture}
            \draw (0, 0) node[inner sep=0]{\includegraphics[width=0.44\textwidth]{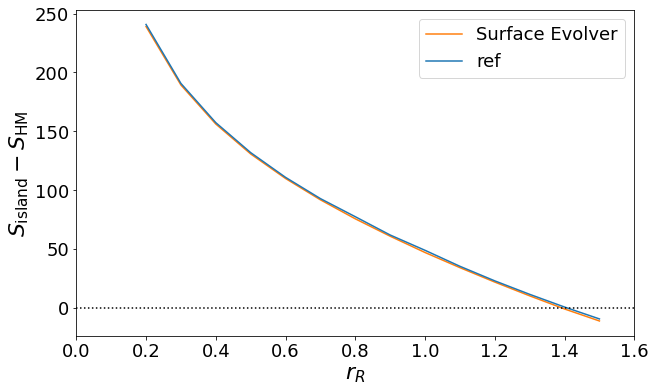}};
            \draw (2.07, 1.4) node[font=\tiny] {\cite{Uhlemann:2021nhu}};
        \end{tikzpicture}
    }
    \hskip 8mm
    \subfigure[][]{\label{fig:del0diff}
        \begin{tikzpicture}
            \draw (0, 0) node[inner sep=0]{\includegraphics[width=0.44\textwidth]{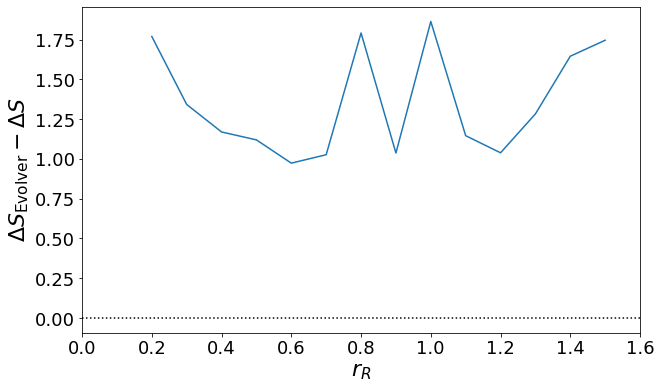}};
            \draw (-3.2, 1.12) node[rotate=90] {\scalebox{0.46}{\cite{Uhlemann:2021nhu}}};
        \end{tikzpicture}
    }
    \caption{Left: Comparison of $\Delta S = S_\mathrm{island}-S_\mathrm{HM}$ as computed by the Surface Evolver and \cite{Uhlemann:2021nhu}. Here, $N_5/K=1.6$. Right: Difference in  $\Delta S$ as computed by the Surface Evolver versus \cite{Uhlemann:2021nhu}.}
\end{figure}

\medskip
\textbf{Match to $\delta=0$ results:}
Our setup reduces to the case studied in \cite{Uhlemann:2021nhu} when the brane separation $\delta$ is taken to zero. Since the surface minimization algorithm used for this paper differs from that of \cite{Uhlemann:2021nhu}, we now present a comparison of the results. This offers an important consistency check, since the algorithms differ significantly in the method of discretization, implementation of boundary conditions, and cutoffs.
Fig. \ref{fig:del0comp} shows the area difference between the island and HM surface for $N_5/K=1.6$ as computed by both the Surface Evolver and the implementation of \cite{Uhlemann:2021nhu}. This quantity is cutoff-independent and should match between the two methods. Despite the differences in implementation, the curves are barely distinguishable. As shown in fig. \ref{fig:del0diff}, the absolute discrepancy between the two methods is at the per cent level compared to the overall scale of the area difference. 

\medskip
\textbf{Parameter space of minimal surfaces:}
To map out the regions in ($\delta$, $N_5/K$, $r_R$) parameter space where each type of surface exists, we manually search out a single choice of parameters where the Surface Evolver algorithm successfully produces a minimal surface. We then increment one parameter in small steps until it fails to do so. For HM surfaces, failure occurs during the evolution when the surface migrates all the way until it hits the poles, causing an error or a spontaneous jump to the partial island type. Partial islands fail to exist either by receding back to the first set of poles or crossing over the second set. In the course of our numerical study, full islands never failed to exist at any point in the parameter space. This result is consistent with the $\delta=0$ case in \cite{Uhlemann:2021nhu}.

\bibliographystyle{JHEP}
\bibliography{islands}
\end{document}